\documentclass[aps,prd,nofootinbib,longbibliography,superscriptaddress,floatfix]{revtex4-2}

\usepackage{graphicx}
\usepackage{amsmath,amssymb,mathtools,bm}
\usepackage{booktabs}
\usepackage{array}
\usepackage{multirow}
\usepackage{tabularx}
\usepackage{microtype}
\usepackage{placeins}
\usepackage{xcolor}
\usepackage[colorlinks=true,linkcolor=blue,citecolor=blue,urlcolor=blue]{hyperref}
\graphicspath{{figures/}}

\newcommand{\SGL}{\mathrm{SGL}}
\newcommand{\SNR}{\mathrm{SNR}}
\newcommand{\SNRc}{\mathrm{SNR}_{C}}
\newcommand{\SNRr}{\mathrm{SNR}_{R}}
\newcommand{\AU}{\mathrm{AU}}
\newcommand{\pc}{\mathrm{pc}}
\newcommand{\m}{\mathrm{m}}
\newcommand{\km}{\mathrm{km}}
\newcommand{\s}{\mathrm{s}}
\newcommand{\um}{\mu\mathrm{m}}
\newcommand{\nas}{\mathrm{nas}}
\newcommand{\data}{\bm{y}}
\newcommand{\truth}{\bm{O}}

\newcommand{\op}{\bm{H}}
\newcommand{\noise}{\bm{n}}
\newcommand{\bkg}{\bm{b}}

\newcommand{\T}{\mathrm{T}}
\newcommand{\FRC}{\mathrm{FRC}}
\newcommand{\SSIM}{\mathrm{SSIM}}
\newcommand{\SBR}{\mathrm{SBR}}

\newcolumntype{Y}{>{\raggedright\arraybackslash}X}

\begin{document}

\title{Direct High-Resolution Imaging of Earth-like Exoplanets \\ with the Solar Gravitational Lens}

\author{Slava G. Turyshev}
\affiliation{ 
Jet Propulsion Laboratory, California Institute of Technology,\\
4800 Oak Grove Drive, Pasadena, CA 91109-0899, USA
}%

\date{\today}

\begin{abstract}
We present a scalar, aperture-averaged observability benchmark for resolved exoplanet imaging with the solar gravitational lens (SGL).  The model uses the source-to-image-plane compression $\bm{x}\simeq-(z/z_0)\bm{x}'$, the finite-aperture kernel $K(\rho>0)\simeq d/(4\rho)$, and an Einstein-ring photon-count model to propagate a real Earth luminance map through a scanned SGL image plane.  Photon noise, finite-exposure smear, cloud-like temporal variability, structured coronal and detector-calibration residuals, optical-operator mismatch, and navigation error are injected as controlled perturbations and reconstructed with a regularized Fourier/Wiener inverse.  For an Earth-radius planet at $30\,\pc$ observed from $650\,\AU$ with a $1\,\m$ telescope, the image cylinder is $1.338\,\km$ across; a $128\times128$ raster has $10.46\,\m$ image-plane pitch, $99.55\,\km$ source pixels, $Q_{\rm exo}=8.01\times10^4\,\s^{-1}$, $Q_{\rm cor}=6.20\times10^9\,\s^{-1}$, and $\SNRc=43.16$ for $1800\,\s$ dwell per sample. Under calibrated scalar ring extraction and the stated temporal, background, calibration, and navigation mitigations, C8 gives reconstructed-map signal-to-noise statistic $\SNRr=6.89$, structural similarity index $\SSIM=0.848$, normalized root-mean-square error (NRMSE) $=0.439$, and a $232\,\km$ Fourier-ring-correlation resolution proxy; a higher-count C9 case gives $\SNRr=10.49$, $\SSIM=0.927$, NRMSE $=0.287$, and $\FRC_{50}=199\,\km$, where the latter value is the two-pixel grid floor of the adopted raster rather than an independent half-data resolution measurement.  Controlled cloud branches show that a static inverse is inappropriate for rotating cloudy data, while phase-registered robust coadds recover persistent surface-proxy information for both advected-proxy (C3b) and independent stochastic cloud-field (C3c) stress tests.  The dominant modeled limitations are temporal sampling, coronal and detector calibration, calibrated ring extraction, image-plane metrology, optical-operator knowledge, and dynamic inversion, rather than calibrated monopole SGL blur.  Thus the benchmark demonstrates recoverable $200$--$230\,\km$-class broadband spatial contrast in the scalar model and converts the SGL concept into quantitative observing, calibration, and inversion requirements for mapping targets at tens of parsecs; it does not by itself validate propulsion, communications, coronagraphic or external-occulter propagation, physical solar multipoles, physical cloud fields, or a full dynamic retrieval pipeline.

\end{abstract}

\maketitle

\section{Introduction}
\label{sec:intro}

The direct imaging of an Earth-like exoplanet is limited simultaneously by angular resolution, contrast, photon statistics, and inference.  A terrestrial planet with radius $R_\oplus$ at $10\,\pc$ subtends only $8.5\,\mu\mathrm{as}$ in diameter.  A coarse $10\times10$ surface map therefore requires resolution elements near $0.85\,\mu\mathrm{as}$; at $\lambda=550\,\mathrm{nm}$, this implies a filled-aperture diameter of order $160\,\km$ or an interferometric baseline of order $130\,\km$ before allowing for stellar suppression, finite throughput, wavefront stability, calibration overhead, exozodiacal light, detector noise, or planetary variability.  A detailed assessment of direct high-resolution exoplanet imaging shows that this surface-resolution benchmark is beyond conventional remote architectures under credible mission assumptions \cite{Turyshev2026exoimage}.  Present and proposed coronagraphic and starshade systems are powerful for detection and disk-integrated spectroscopy, but they remain limited by telescope diffraction, contrast stability, throughput, and astrophysical backgrounds \cite{Traub2010,Galicher2023,LUVOIR2019,HabEx2020,Cash2006}.  Interferometric concepts move the angular-resolution burden to long-baseline formation flying, optical-path stability, high-contrast suppression, and dense Fourier-plane sampling \cite{Bracewell1978,Angel1997,Labeyrie1996}.  Indirect photometric mapping and spin-orbit tomography can recover low-order longitudinal or albedo structure, but they do not replace direct surface-resolved imaging \cite{CowanAgol2008,FujiiKawahara2012,CowanFujii2018}.

The solar gravitational lens (SGL) changes this scaling because the Sun's gravitational field supplies the dominant optical element.  The idea follows from Einstein's gravitational lensing argument and Eshleman's proposal to exploit the solar gravitational focus \cite{Einstein1936,Eshleman1979}.  Modern wave-optical treatments show that Maxwell's equations on the post-Newtonian solar metric produce a strong-interference region beginning beyond $R_\odot^2/(2r_g)\simeq547.8\,\AU$, where $r_g=2GM_\odot/c^2=2.95\,\km$ is the solar Schwarzschild radius \cite{Turyshev2017PRD95,TuryshevToth2017PRD96}.  At optical and near-infrared wavelengths, the ideal monopole gain is of order $4\pi^2r_g/\lambda$ and the angular response is of order $0.38\lambda/b$, corresponding to nanoarcsecond and sub-nanoarcsecond angular scales for solar-grazing rays.  For an extended source, however, the SGL does not form a conventional focal-plane image.  It maps the exoplanet into a narrow image cylinder in the SGL focal region; a spacecraft must traverse this image plane, measure the Einstein-ring flux around the Sun at each sampling location, and reconstruct the source by solving an inverse problem \cite{TuryshevToth2020Photometric,TuryshevToth2020Process,TuryshevToth2020Extended,TothTuryshev2021Recovery,TuryshevToth2022Resolved}.

The SGL observing mode is intrinsically an image-plane sampling and inversion problem.  A spacecraft in the SGL focal region samples the kilometer-scale projected image cylinder, integrates the Einstein-ring flux at many image-plane locations, time-tags those measurements, and reconstructs the source by solving an inverse problem.  Spacecraft motion, integration time, sampling cadence, calibration, and regularized inversion are therefore part of the instrument model.  The same optical gain and angular response that make the SGL unique also impose specific requirements on solar-corona photon noise, solar multipole perturbations, the broad $1/\rho$ aperture-averaged point-spread function, host-star leakage, detector calibration against a bright background, finite-exposure smear, target rotation, cloud evolution, image-plane navigation, pointing jitter, and reconstruction bias.  The relevant technical framework combines wave optics, gravitational lensing, coronagraphy and external occultation, detector photon-transfer modeling, astrodynamics, and statistical inverse problems.

The most useful first-order reconstruction algorithms are Fourier quotient and Wiener/Tikhonov methods, because they provide transparent baselines and expose the conditioning of the SGL optical transfer function \cite{Goodman2005,Wiener1949,TikhonovArsenin1977,BerteroBoccacci1998,Hansen2010}.  For photon-dominated data, Richardson--Lucy and other Poisson-likelihood methods are natural alternatives \cite{Richardson1972,Lucy1974}.  For a flight-quality pipeline, the reconstruction must become a joint Bayesian or variational dynamic inverse problem in which the planet map, cloud field, point spread function (PSF), coronal residuals, detector calibration, ephemeris, and pointing state are inferred together \cite{KaipioSomersalo2005,Tarantola2005,BarrettMyers2004,VanTrees2001,Kay1993}, see Table~\ref{tab:notation} for notation and abbreviations used throughout.

This paper develops and applies a fully specified numerical benchmark for SGL imaging with a real Earth input image.  It defines a physically explicit simulation framework whose assumptions, operators, noise terms, reconstruction method, validation metrics, and case definitions are stated transparently. The simulation begins with a NASA Apollo 17 Earth image, converts it to a normalized scalar luminance test map, propagates it through an aperture-averaged SGL operator, injects individual noise and systematic terms, reconstructs the source with a non-oracle regularized inverse, and quantifies the image degradation introduced by each modeled effect.  The model separates effects that are physically well established---the monopole optical response, image-plane compression, solar-corona photon statistics, finite-aperture averaging, and Fourier deconvolution---from controlled surrogates used to test sensitivity to solar-multipole mismatch, host-star leakage, cloud-like temporal variability, and image-plane navigation warp.

The objective of this paper is to convert realistic degradations into first-order quantitative observing and reconstruction requirements.  In the scalar aperture-integrated model studied here, calibrated monopole SGL blur is not the dominant limitation; the limiting terms are temporal sampling, cloud and rotation modeling, coronal and detector calibration residuals, navigation error, calibrated ring extraction, and imperfect knowledge of the optical operator.  This distinction is central to the mission case.  The SGL supplies angular resolution and optical gain that conventional remote architectures do not provide, but that information is useful only if the observatory architecture preserves it through propulsion to the focal region, solar-background suppression, calibrated ring extraction, image-plane metrology, phase-registered observing, communications, and statistically controlled inversion.  The relevant question is therefore not whether the SGL is a valid optical concept, but which mission/inference requirements must be met to exploit it.

The claim tested in this paper is deliberately narrow.  Given the stated scalar, aperture-averaged SGL operator, an assumed calibrated ring-extraction statistic, specified noise and nuisance terms, and a non-oracle regularized inverse, the benchmark asks whether broadband spatial contrast from an Earth-like test scene remains recoverable and what observing, calibration, metrology, and inversion requirements control that recovery.  The paper does not claim end-to-end mission feasibility or physical validation of individual recovered surface features.  In particular, it does not validate propulsion to the focal region, communications, physical coronagraphic or external-occulter propagation, ring-sector or spectral extraction, physical solar-multipole diffraction, physical cloud radiative transfer, or a full spin-resolved dynamic retrieval.  Mission-level validation requires a wave-optical SGL+telescope PSF and ring-response library, wavelength-dependent time-tagged Earth or GCM+radiative-transfer scenes, an end-to-end calibration and metrology budget, and a dynamic inverse problem that estimates the planet map and nuisance parameters jointly.  

The paper is organized as follows.  Section~\ref{sec:science_cases} defines the target classes and scientific data products.  Section~\ref{sec:physics} summarizes SGL image formation and the optical response.  Section~\ref{sec:framework} presents the simulation framework, input data, and model-selection strategy.  Section~\ref{sec:forward} gives the forward model, including noise, detector, temporal, background, multipole, navigation, and pointing effects.  Section~\ref{sec:reconstruction} describes the reconstruction and regularization method.  Section~\ref{sec:validation} defines validation metrics and tests.  Section~\ref{sec:results} presents the stepwise simulation results.  Section~\ref{sec:requirements} derives sensitivity and requirements.  Section~\ref{sec:architectures} discusses observing strategies and mission architectures.  Section~\ref{sec:comparison} compares the SGL with conventional exoplanet-imaging approaches.  Section~\ref{sec:limitations} identifies limitations and next steps, and Sec.~\ref{sec:conclusions} summarizes the conclusions.

\begin{table}[h!]
\centering
\caption{Notation and abbreviations used throughout the exoplanet-imaging benchmark.  The convention $n$ for the linear raster dimension and $N=n^2$ for the total number of image-plane samples is used throughout.}
\label{tab:notation}
\begin{tabular}{ll}
\toprule
Symbol or abbreviation & Meaning \\
\midrule
SGL & solar gravitational lens \\
PSF & point-spread function \\
SNR & signal-to-noise ratio \\
SBR & source-to-background ratio, $\SBR=Q_s/Q_b$; for the exo-Earth case $\SBR_{\rm exo}=Q_{\rm exo}/Q_{\rm cor}$ \\
FRC & Fourier-ring correlation; $\FRC_{50}$ is the 50\% resolution proxy \\
FOV & field of view, when referring to a conventional instrument or selected SGL subfield \\
AU, pc & astronomical unit and parsec \\
$z$, $z_0$ & heliocentric observer distance and target distance \\
$\boldsymbol{\rho}$, $\boldsymbol{\xi}$ & SGL image-plane and source-plane coordinates; $\boldsymbol{\rho}=-(z/z_0)\boldsymbol{\xi}$ \\
$D_{\rm img}$ & projected image-cylinder diameter \\
$n$, $N$ & linear raster dimension and total sample count, $N=n^2$ \\
$\Delta_{\rm img}$ & image-plane pitch, $D_{\rm img}/n$ \\
$d$ & telescope diameter \\
$K(\rho)$ & aperture-averaged scalar SGL kernel \\
$Q_{\rm exo}$, $Q_{\rm cor}$ & exoplanet signal rate and solar-corona background rate in the fiducial scalar benchmark \\
$\SNRc$, $\SNRr$ & convolved-raster SNR and post-reconstruction residual SNR statistic \\
\bottomrule
\end{tabular}
\end{table}

\section{Scientific use cases and target classes}
\label{sec:science_cases}

The primary scientific return of an SGL imaging mission is a hierarchy of calibrated, resolved maps rather than a single deblurred picture.  At the measurement level the observatory records time-tagged Einstein-ring fluxes at known image-plane coordinates and spacecraft states.  The principal imaging products are: (i) persistent broadband albedo and surface maps; (ii) cloud occurrence and variability statistics; (iii) rotationally resolved surface--atmosphere maps; and (iv) regional temporal monitoring of evolving features.  Disk-integrated spectroscopy, spectropolarimetry, and thermal channels are natural extensions of the same image-plane data model, but in this paper they are treated only as future supporting diagnostics for interpreting resolved maps, not as deliverables of the benchmark.  A static broadband map is therefore the validation product for the scalar benchmark, while the eventual science product is a dynamic, calibrated map with uncertainty estimates.

Two imaging modes organize the benchmark.  Broadband image-plane rastering measures the total Einstein-ring flux in a reflected-light band at many transverse spacecraft positions and is the appropriate mode for albedo, cloud, and surface-morphology maps.  Ring-sector photometry retains azimuthal information around the Einstein ring and can provide low-order spatial constraints during a dwell, but it requires a physical coronagraph or external-occulter propagation model.  Selected wavelength-resolved or polarimetric measurements can be added to the same raster or to a small number of high-value regions; here their role is limited to improving interpretation of the recovered map, for example by distinguishing persistent surface albedo from transient atmospheric structure.

The SGL is uniquely valuable when both angular resolution and photons are limiting, and it should be assessed as an in-situ scanning observatory rather than as an unrealistically large conventional aperture.  A conventional coronagraph or starshade can suppress the host star but cannot remove the telescope diffraction requirement; a filled aperture would need a diameter of order $10^2\,\km$ to resolve an Earth analog at $10\,\pc$ into even a coarse map.  A free-flying interferometer can provide angular resolution in principle, but must also supply collecting area, contrast suppression, path-length stability, and dense Fourier coverage.  By contrast, the SGL moves the angular-resolution problem into a scanned image-plane measurement: the telescope diameter controls ring collection and extraction, while the source-plane resolution is set by image-plane sampling, temporal registration, and the inverse problem.  This trade does not make the mission easy, but it converts an otherwise prohibitive aperture requirement into propulsion, metrology, calibration, communications, and inference requirements that can be allocated and tested.

\begin{figure}[t]
\centering
\includegraphics[width=0.82\textwidth]{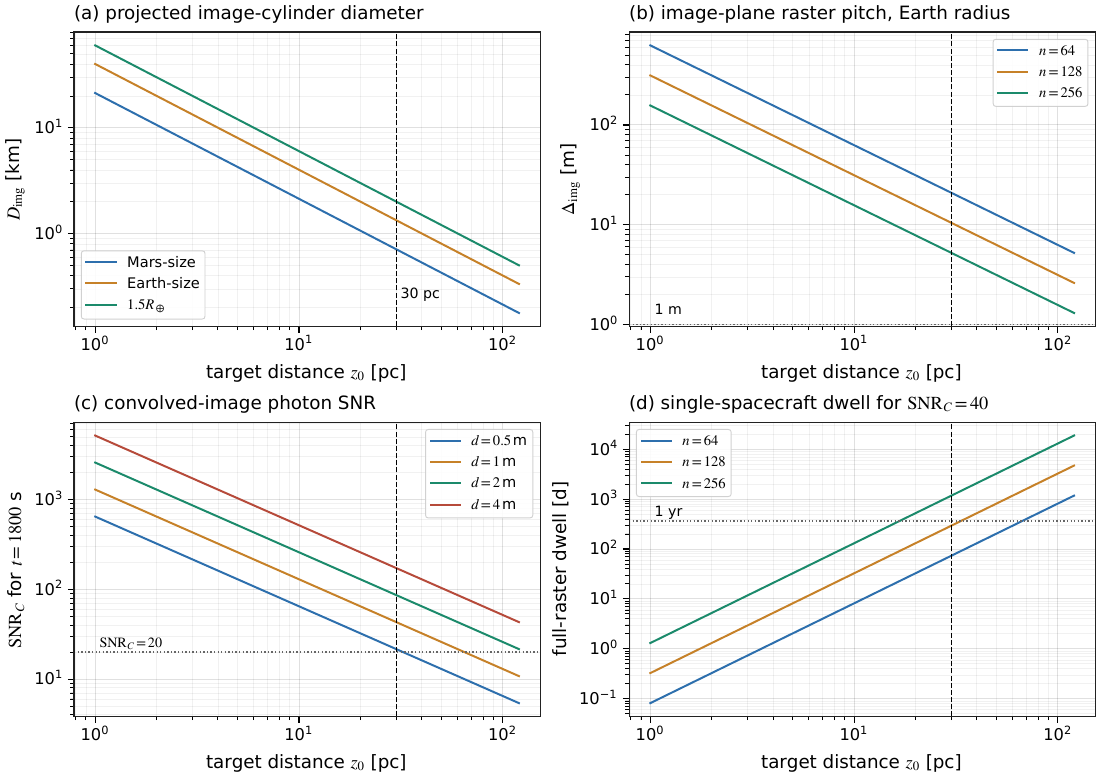}
\caption{Target-distance and sampling scalings for SGL imaging.  Panel (a) shows the projected image-cylinder diameter for representative planet radii.  Panel (b) shows the image-plane pitch for an Earth-radius planet as a function of linear grid size.  Panel (c) shows the pre-deconvolution convolved-image SNR for $t=1800\,\s$ per image-plane sample.  Panel (d) gives the full-raster dwell required to reach $\SNRc=40$ with one spacecraft.  The dashed vertical line marks the fiducial $30\,\pc$ benchmark; the horizontal guides mark practical scale references.}
\label{fig:target_scaling}
\end{figure}

The most favorable targets are nearby planets for which the orbit, ephemeris, radius, phase function, and preferably rotation period are known before the spacecraft reaches the focal region.  For a planet of radius $R_p$, the image-cylinder diameter and raster pitch follow directly from the standard SGL image-compression geometry \cite{TuryshevToth2020Photometric,TuryshevToth2020Extended,TothTuryshev2021Recovery,TuryshevToth2022Resolved} and are
\begin{align}
D_{\rm img} &= 1.338
\left(\frac{R_p}{R_\oplus}\right)
\left(\frac{z}{650\,\AU}\right)
\left(\frac{30\,\pc}{z_0}\right)\,\km, \\
\Delta_{\rm img} &= \frac{D_{\rm img} }{n}=10.46
\left(\frac{128}{n}\right)
\left(\frac{R_p}{R_\oplus}\right)
\left(\frac{z}{650\,\AU}\right)
\left(\frac{30\,\pc}{z_0}\right)\,\m .
\label{eq:target_scaling_delta}
\end{align}
Thus more distant targets are not harder because the desired source-pixel scale changes; for fixed $n$ the source scale is $2R_p/n$.  They are harder because the projected image cylinder and the image-plane pitch shrink, the planet photon rate falls approximately as $z_0^{-1}$ in the adopted broadband scaling, and the dwell time required to reach a given $\SNRc$ grows approximately as $z_0^2$ in the corona-dominated limit.  The dwell per sample required for a target convolved-image SNR is
\begin{equation}
 t(\SNRc)=\SNRc^2\frac{Q_{\rm exo}+Q_{\rm cor}}{Q_{\rm exo}^2},
 \qquad
 T_{\rm dwell}=\frac{n^2 t(\SNRc)}{N_{\rm sc}},
\label{eq:dwell_for_snr}
\end{equation}
where $N_{\rm sc}$ is the number of independent spacecraft or equivalent interleaved sampling platforms.  Figure~\ref{fig:target_scaling} illustrates these scalings for the fiducial optical model and shows why distance, aperture, grid size, and spacecraft multiplicity are coupled design variables.

The target classes that best match these requirements are nearby terrestrial planets around quiet stars, planets with previously measured rotation periods or repeated photometric phase curves, and systems in which the host-star, exozodiacal, and ephemeris environment can be modeled before arrival.  Earth analogs and super-Earths are the prime targets because the SGL can connect global context to regional albedo, clouds, and persistent surface morphology at $\sim 100$--$300\,\km$ scales.  Giant planets and moon systems are useful secondary imaging targets because belts, storms, rings, and thermal structure are bright, extended, and dynamically informative.  Less favorable targets are not excluded, but they move the problem from static mapping toward a joint dynamic inverse problem in which planetary variability, background leakage, and geometric parameters must be estimated simultaneously.

The controlling terms for these products are introduced in the model sections rather than summarized in a separate roadmap graphic: photon statistics and calibration control the broadband map, phase coverage controls cloud and rotation products, and navigation plus dynamic inversion control regional monitoring.  Clouds are therefore not treated only as contaminants.  They are nuisance structure for a persistent surface map, but their occurrence, morphology, and temporal correlations are also atmospheric and climate observables when the data are modeled as a time series.  This perspective keeps the benchmark focused on quantitative image recovery while preserving the broader scientific value of repeated SGL observations.

\section{Physical model of SGL imaging}
\label{sec:physics}

\subsection{Image-plane compression}

For an extended source at distance $z_0$, the monopole SGL maps a source-plane coordinate $\boldsymbol{\xi}\equiv\bm{x}'$ into an image-plane coordinate $\boldsymbol{\rho}\equiv\bm{x}$ according to the standard image-compression relation derived in Refs.~\cite{TuryshevToth2020Photometric,TuryshevToth2020Process,TuryshevToth2022Resolved},
\begin{equation}
\boldsymbol{\rho}\equiv\bm{x}\simeq
-\frac{z}{z_0}\boldsymbol{\xi}
\equiv -\frac{z}{z_0}\bm{x}' ,
\label{eq:mapping}
\end{equation}
where $z$ is the heliocentric distance of the observer.  The minus sign denotes the parity reversal of the SGL image cylinder.  This relation is the geometric link between the source-plane map and the image-plane scan; it is the basis for the image-cylinder diameter, raster pitch, and navigation requirements used throughout the paper.  An Earth-radius planet is therefore compressed into an image cylinder of diameter
\begin{equation}
D_{\rm img}=2R_\oplus \frac{z}{z_0}
=1.338\left(\frac{z}{650\,\AU}\right)\left(\frac{30\,\pc}{z_0}\right)\km .
\label{eq:Dimg}
\end{equation}
For $n$ linear source pixels, the image-plane sample pitch is
\begin{equation}
\Delta_{\rm img}=\frac{D_{\rm img}}{n}.
\end{equation}
In the fiducial $30\,\pc$, $650\,\AU$, $n=128$ case, $\Delta_{\rm img}=10.46\,\m$.  This spacing is more than ten telescope diameters for a $1\,\m$ aperture, a fact that strongly mitigates the deconvolution penalty.

\subsection{SGL point-spread function and aperture averaging}

The monopole SGL amplification factor for a point source contains the Bessel term \cite{Turyshev2017PRD95,TuryshevToth2017PRD96}
\begin{equation}
\mu_{\SGL}(\rho)=\mu_0 J_0^2\!\Big(k\rho\sqrt{\frac{2r_g}{z}}\Big),\qquad
\mu_0\simeq 4\pi^2\frac{r_g}{\lambda},
\label{eq:psf_bessel}
\end{equation}
where $k=2\pi/\lambda$ is the optical wavenumber, $\rho$ is the transverse separation in the SGL image plane, and $J_0$ is the Bessel function of the first kind.  The quantity $\mu_0$ denotes the ideal on-axis monopole gain.

The raw Bessel oscillations have centimeter-scale structure in the image plane at optical wavelengths.  In the present work this exact point-source expression is used only to anchor the transition to the aperture-integrated SGL kernel employed in the numerical imaging benchmark; the actual reconstruction calculations use Eq.~(\ref{eq:kernel}), not a direct inversion of Eq.~(\ref{eq:psf_bessel}).  A meter-class telescope used as a photometric ring collector averages over many oscillations.  The finite-aperture, aperture-averaged kernel is well approximated off-axis by a long tail proportional to $d/(4\rho)$ \cite{TothTuryshev2021Recovery,TuryshevToth2020Extended}.  The implemented scalar kernel is
\begin{equation}
K(0)=1,\qquad K(\rho>0)\simeq \frac{d}{4\rho},
\label{eq:kernel}
\end{equation}
renormalized so that the mean convolved disk signal is unity.  This approximation defines the scalar aperture-integrated operator used in the scanned image-plane raster; full telescope focal-plane propagation enters through the ring-extraction model discussed in Sec.~\ref{sec:physics}.

\subsection{Einstein-ring photometry}

A spacecraft in the SGL focal region looks back toward the Sun and observes the planet's light as an Einstein ring.  The ring angular radius is
\begin{equation}
\theta_{\rm ER}=\sqrt{\frac{2r_g}{z}},
\end{equation}
while the solar angular radius is $R_\odot/z$.  At $z=650\,\AU$, a telescope with $f=10\,\m$ places the solar limb and Einstein ring near radii of $71.6\,\mu\m$ and $77.9\,\mu\m$ on the focal plane, respectively.  The scalar simulation integrates the ring into one photometric measurement per image-plane sample.  This is conservative relative to a future ring-segment inversion, because the full focal-plane ring contains azimuthal information about the source distribution \cite{TuryshevToth2020Process,TuryshevToth2022Spectral}.

The scalar annular measurement used in this paper assumes that the ring-extraction problem has already been reduced to an effective ring-extraction throughput.  This defines the optical-extraction approximation used by the benchmark.  Eq.~\eqref{eq:ring_limb_separation} is simply the small-angle projection of the solar limb and Einstein ring into an instrument focal plane \cite{TuryshevToth2020Process,TuryshevToth2022Spectral}: $f$ is the effective focal length, $R_\odot$ is the solar radius, $z$ is the heliocentric observer distance, $r_g$ is the solar Schwarzschild radius, $r_\odot$ is the solar-limb radius on the detector, $r_{\rm ER}$ is the Einstein-ring radius, and $\Delta r_{\rm ER-\odot}$ is their radial separation.  Thus
\begin{align}
 r_\odot(f,z) &= f\frac{R_\odot}{z}, \\
 r_{\rm ER}(f,z) &= f\sqrt{\frac{2r_g}{z}}, \\
 \Delta r_{\rm ER-\odot} &= f\Big(\sqrt{\frac{2r_g}{z}}-\frac{R_\odot}{z}\Big).
\label{eq:ring_limb_separation}
\end{align}
At $z=650\,\AU$ and $f=10\,\m$, $r_\odot=71.6\,\mu\m$ and $r_{\rm ER}=77.9\,\mu\m$, so the radial separation is only $\Delta r_{\rm ER-\odot}\simeq6.3\,\mu\m$.  For a $d=1\,\m$ telescope at $\lambda=1\,\um$, the Airy radius is $1.22\lambda f/d=12.2\,\mu\m$; the ring-limb separation is therefore only $0.52$ Airy radii.  Consequently, an internal coronagraph or focal-plane mask cannot be represented by a scalar annular aperture without an explicit diffraction and throughput calculation.  In this paper, annular aperture-integrated photometry therefore assumes an effective ring-extraction throughput and an already defined background-subtraction statistic.  It does not demonstrate coronagraphic suppression, external-occulter performance, coronal subtraction, or leakage calibration.  A mission-level forward model must include an internal coronagraph or external occulter propagation term, $\eta_{\rm ring}(\lambda,z,d,f)$, wavelength-dependent ring throughput, solar-corona leakage, host-star and exozodiacal backgrounds, and the corresponding covariance in the measurement model.

\begin{figure}[t]
\centering
\includegraphics[width=0.82\textwidth]{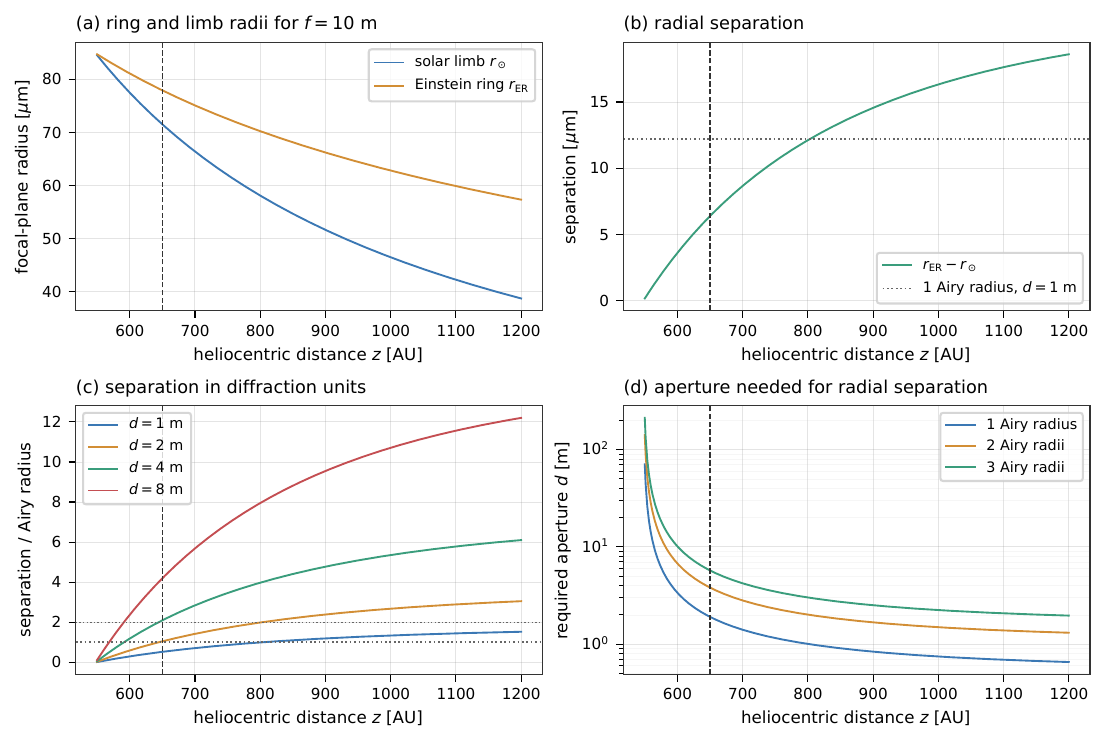}
\caption{Einstein-ring extraction is an optical-propagation problem.  Panel (a) shows the solar-limb and Einstein-ring radii in a $f=10\,\m$ focal plane.  Panel (b) compares their radial separation with the $d=1\,\m$ Airy radius.  Panel (c) expresses the separation in diffraction units for several telescope apertures.  Panel (d) gives the aperture required for one, two, or three Airy radii of radial separation.  The fiducial $650\,\AU$ case is marginal for a $1\,\m$ internal-coronagraph aperture-integrated measurement and sets an occulter/coronagraph design requirement for converting the ring into a calibrated scalar photometric observable.  The scalar annular model assumes an effective ring-extraction throughput and an already calibrated photometric statistic; diffraction propagation, coronal subtraction, and leakage covariance enter the mission-level forward model.}
\label{fig:ring_limb_separation}
\end{figure}

Figure~\ref{fig:ring_limb_separation} summarizes this ring--limb separation in focal-plane and diffraction units and shows why calibrated ring extraction is an optical-design requirement rather than an assumption that can be hidden inside the scalar convolution.

\subsection{Solar multipoles}

A realistic Sun is not a perfect monopole.  Zonal harmonics and rotation perturb the gravitational phase \cite{TuryshevToth2021Extended,TuryshevToth2021ExtendedGL,TuryshevToth2021Distribution,TuryshevToth2022Multipole}.  Schematically,
\begin{equation}
\Phi(\bm{b})=-2kr_g\ln(kb)+kr_g\sum_{\ell\ge 2}J_\ell\Big(\frac{R_\odot}{b}\Big)^\ell F_\ell(\phi_b,\theta_\odot),
\label{eq:multipole_phase}
\end{equation}
where $\bm{b}$ is the impact-parameter vector and $F_\ell$ encodes angular dependence.  The quadrupole produces astroid caustics and can change the focal-plane structure from a circular ring into arcs or multiple bright spots for point sources \cite{TuryshevToth2021ExtendedSun}.  For a resolved exoplanet whose projected image is much larger than the caustic, the total flux is not expected to change as strongly as the reconstruction kernel.  The principal error is therefore PSF mismatch: the wrong inverse operator can create false structure or reduce recovered contrast.

The present paper implements a controlled anisotropic PSF-mismatch surrogate rather than a full solar-multipole diffraction calculation; this surrogate is used only to test inverse sensitivity and is not evidence that the physical solar multipoles are negligible:
\begin{equation}
K_{\rm mp}(\rho,\theta)=K(\rho)
\Big(1+\epsilon_2\cos 2(\theta-\theta_q)
        +\epsilon_4\cos 4(\theta-\theta_q)\Big),
\label{eq:multipole_surrogate}
\end{equation}
where $(\rho,\theta)$ are polar coordinates in the SGL image plane and $\theta_q$ sets the orientation of the imposed anisotropy.  The coefficients $\epsilon_2$ and $\epsilon_4$ are dimensionless fractional perturbation amplitudes of the effective aperture-averaged kernel.  The values used here, $\epsilon_2=0.03$ and $\epsilon_4=0.01$, impose $3\%$ and $1\%$ angular
modulations of the kernel tail with azimuthal orders $m=2$ and $m=4$, respectively.  They are not solar gravitational moments, are not computed from $J_2,J_4,\ldots$, and should be interpreted only as controlled sensitivity parameters used to quantify reconstruction sensitivity to anisotropic PSF mismatch.  After applying the modulation, the discrete kernel is renormalized,
\[
K_{\rm mp}\leftarrow K_{\rm mp}
\frac{\sum_{ij}K_{ij}}{\sum_{ij}(K_{\rm mp})_{ij}},
\]
so that the total kernel weight equals that of the monopole kernel.  As stated in the scope paragraph of Sec.~\ref{sec:intro}, Eq.~\eqref{eq:multipole_surrogate} is a scalar residual-sensitivity test.  It probes the effect of an energy-conserving, few-percent anisotropic error in the aperture-averaged operator; it is not a prediction of the wave-optical response of the real solar quadrupole, higher moments, oblateness, plasma, or time-dependent solar figure.

The corresponding mission-level input is a physical SGL+telescope PSF and ring-response library,
\[
K_{\rm phys}=K_{\rm SGL}^{\rm phys}
\!\left(\bm{x},\bm{x}';
\lambda,b,z,J_2,J_4,\ldots,\hat{\bm{s}}_\odot,
\mathcal{P}_\odot(t),\bm{x}_{\rm sc}(t),\bm{x}_{\rm tar}(t)\right),
\]
computed from measured or literature-based solar moments, solar spin orientation, wavelength, impact parameter, heliocentric distance, solar-plasma state, spacecraft ephemeris, and target ephemeris.  Existing extended-Sun and multipole SGL calculations provide the formal basis for this upgrade \cite{TuryshevToth2021Extended,TuryshevToth2021ExtendedGL,TuryshevToth2021Distribution,TuryshevToth2022Multipole,TuryshevToth2021ExtendedSun,TuryshevToth2023Faint}.  The C1a--C1b entries in Table~\ref{tab:metrics} should therefore be read only as the penalty for this surrogate operator.

\section{Simulation framework}
\label{sec:framework}

\subsection{Model selection and implementation strategy}

The simulation is organized around a hierarchy of models.  The baseline operator is the scalar, aperture-averaged, stationary SGL convolution.  This choice is appropriate when the telescope is used as a photometric collector of the Einstein ring and the image-plane sample spacing is many aperture diameters.  It is also the only level at which the forward and inverse problems can be written in a transparent fast-Fourier-transform (FFT) form suitable for controlled case-by-case sensitivity tests.  Higher-fidelity wave-optical models are the natural next step for the focal-plane ring and solar multipoles.

Each physical effect is implemented either as part of the linear operator, as a time-dependent mapping, or as an additive or multiplicative nuisance term.  Monopole blur and residual jitter are operators.  Rotation and finite exposure are source-to-measurement mappings.  Corona photon noise, read noise, dark current, and shot noise are stochastic terms.  Structured coronal residuals, host-star leakage, exozodiacal light, flat-field residuals, and nonlinearity are nuisance fields.  This separation is useful because it permits controlled case analysis: every case is produced by adding one class of realism, reconstructing with a specified inverse model, and evaluating the resulting penalty against the same real-Earth truth image.

This decomposition defines the traceability path for future mission-level modeling.  The optical module should extend Eq.~\eqref{eq:kernel} to the full SGL-telescope diffraction integral, including solar $J_2,J_4,\ldots$, spin orientation, and ephemeris.  The scene module should extend the single-image model in Eq.~\eqref{eq:luminance} to multi-epoch Earth observations or a general-circulation model coupled to radiative transfer.  The background module should replace the nuisance templates in Eq.~\eqref{eq:forward_model} with wavelength-dependent coronal, host-star, exozodiacal, and coronagraphic or external-occulter propagation terms.  The reconstruction should extend the Wiener baseline in Eq.~\eqref{eq:wiener} to a joint dynamic Bayesian estimator for the map and nuisance parameters.  The present paper provides the transparent baseline and first-order degradations needed to expose the controlling requirements.

\subsection{Input data}

\begin{figure}[t]
\centering
\includegraphics[width=0.78\textwidth]{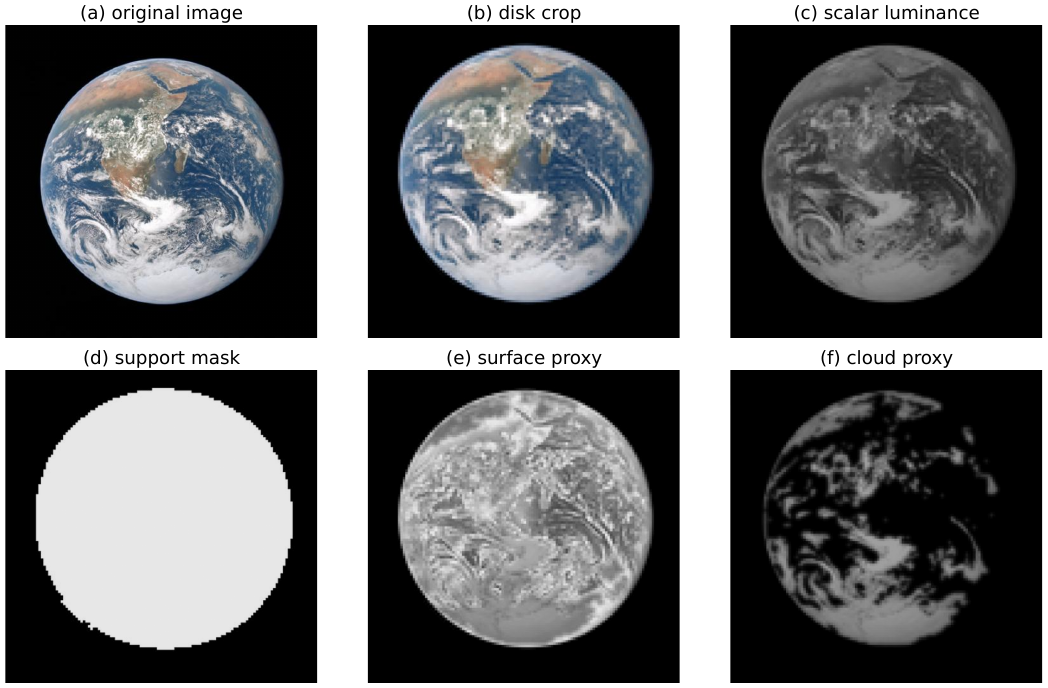}
\caption{Input data and preprocessing used for the scalar benchmark.  Panels (a) and (b) show the original Earth image and the disk crop resampled to the simulation grid.  Panel (c) is the scalar luminance truth, normalized to unit disk mean.  Panel (d) is the reconstruction support mask.  Panel (e) is a cloud-suppressed, surface-like diagnostic proxy produced by attenuating bright, low-saturation cloud-like pixels; panel (f) is the complementary cloud-like diagnostic proxy.  These proxies are used only for temporal sensitivity tests and are not physical surface-reflectance or cloud-optical-depth products.  The figure establishes provenance and preprocessing only; it is not a reconstructed SGL image.}
\label{fig:input}
\end{figure}

The input scene is a real NASA Apollo 17 Blue Marble image \cite{NASABlueMarble}.  The RGB image is cropped around the illuminated disk, resampled to $128\times128$ source pixels, converted to scalar luminance,
\begin{equation}
O(x,y)=0.2126R(x,y)+0.7152G(x,y)+0.0722B(x,y),
\label{eq:luminance}
\end{equation}
and normalized so that the disk mean is unity.  A support mask is obtained by thresholding and retaining the largest connected illuminated component.  The two proxy maps in Fig.~\ref{fig:input} are diagnostic surrogates derived from this same image, not physical retrievals.  The surface-like proxy is a cloud-suppressed grayscale field formed by attenuating the bright low-saturation cloud component and renormalizing the remaining disk intensity.  The cloud-like proxy is the complementary bright, low-saturation component, smoothed and masked to the illuminated disk.  These proxy fields are used only to test sensitivity to temporal incoherence and cloud-like variability, and to define the low-order cloud statistics used by the C3c independent stochastic cloud-field stress test; they do not represent calibrated surface reflectance, cloud optical depth, or a general-circulation-model cloud field.  No synthetic texture is introduced into the baseline input scene.

The normalized luminance image is used only as a morphologically realistic broadband test object.  It is not a calibrated, wavelength-dependent top-of-atmosphere radiance field and does not include a physical bidirectional reflectance model, atmospheric scattering phase function, ocean glint, thermal emission, or time-dependent cloud radiative transfer.  The photon rates in Sec.~\ref{sec:forward} set the measurement noise scale for a fiducial Earth-like source, while the normalized image sets the spatial contrast pattern to be recovered.  Absolute spectrophotometric performance must be evaluated with a wavelength-dependent planet, atmosphere, host-star, and instrument model.

Consequently, the image metrics reported below quantify recognizable spatial-contrast recovery for an analytic/image-based benchmark scene, not a physically validated retrieval of persistent surface properties through realistic cloud fields.  A next-stage Earth-validation data set should use wavelength-dependent, time-tagged top-of-atmosphere Earth observations, or a general-circulation model coupled to radiative transfer, so that cloud optical depth, cloud altitude, surface bidirectional reflectance, ocean glint, atmospheric scattering, thermal emission, viewing geometry, and temporal correlations are generated by a physical scene model rather than by image-derived proxies.  In that validation, the relevant claim would be recovery of persistent surface information with calibrated uncertainty under realistic cloud statistics, not merely recognition of familiar spatial contrast in a single broadband image.

\subsection{Discrete forward model}

The measurement model is
\begin{equation}
\data=\op\truth+\bkg+\noise,
\label{eq:forward_model}
\end{equation}
where $\op$ includes the SGL optical response, temporal smearing, and coordinate mapping; $\bkg$ contains structured residuals from the solar corona, host star, exozodi, and detector calibration; and $\noise$ contains stochastic photon and detector noise.  In the stationary monopole branch, $\op$ is a convolution:
\begin{equation}
 y_{ij}=\sum_{p,q}K(x_i-x_p,y_j-y_q)O_{pq},
\end{equation}
computed by FFT on a $192\times192$ guard-padded grid:
\begin{equation}
\widehat{y}(\bm{f})=\widehat{K}(\bm{f})\widehat{O}(\bm{f}).
\label{eq:fft_convolution}
\end{equation}
The guard band reduces circular-convolution artifacts.  The same FFT operator is used in the inverse model when the forward model is stationary.

\begin{figure}[t!]
\centering
\includegraphics[width=0.82\textwidth]{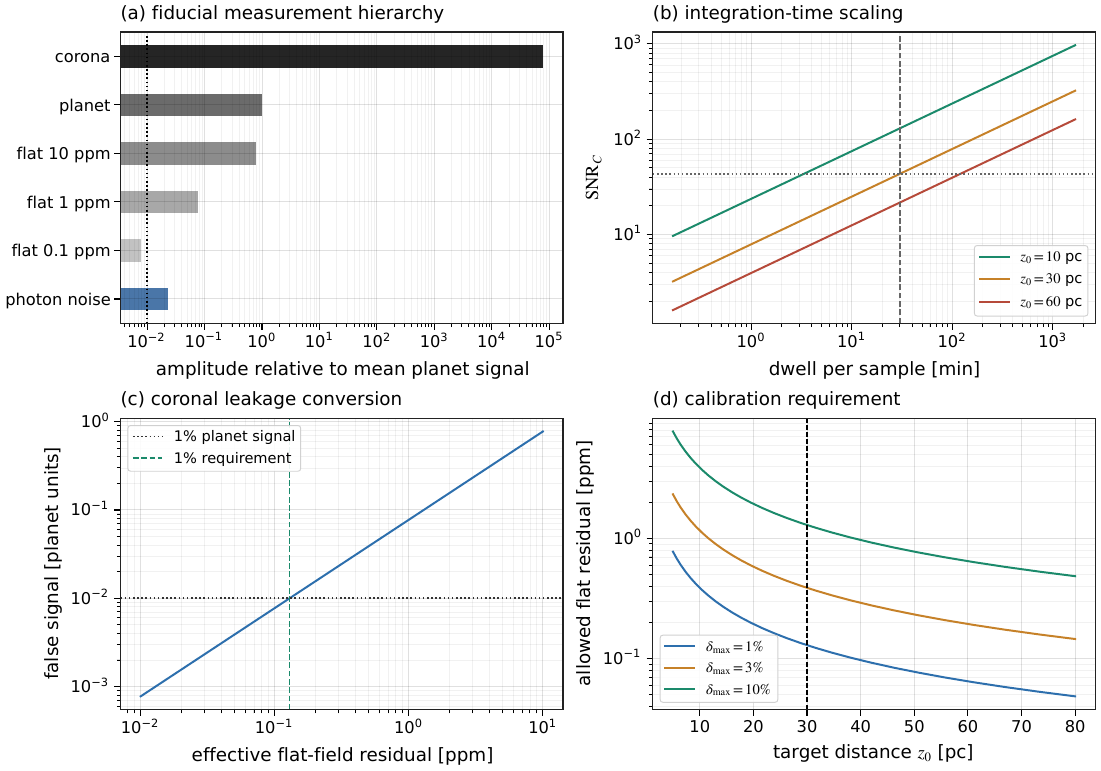}
\caption{SNR and calibration error budget for the fiducial scalar benchmark.  Panel (a) shows the measurement hierarchy in planet units; the corona dominates the raw photon rate, while the planet signal defines unity.  Panel (b) shows the convolved-image SNR growth with dwell time for representative target distances.  Panel (c) converts multiplicative flat-field residuals on the bright coronal signal into false planet-unit signals using $\delta_{\rm planet}=\delta_{\rm flat}Q_{\rm cor}/Q_{\rm exo}$.  Panel (d) gives the effective flat-field residual allowed if coronal calibration leakage is to remain below 1, 3, or 10 percent of the mean planet signal. The horizontal axis should be interpreted as the effective residual after the full ring-extraction, coronal-subtraction, detector-calibration, pointing/metrology, and template-subtraction chain, not as the raw stability of one detector parameter.}
\label{fig:error_budget_control}
\end{figure}

For requirements work it is useful to write the same model in count units before mean-background subtraction.  For sample $k$ and wavelength channel $a$,
\begin{equation}
 y_{ka}=t_{ka}\left[\eta_{p,a}Q_{{\rm exo},a}(\op_{ka}\truth)_k
 +\eta_{{\rm cor},a}Q_{{\rm cor},a}+q_{\star,ka}+q_{{\rm zodi},ka}
 +q_{{\rm dark},ka}\right]+r_{ka}+s_{ka},
\label{eq:count_model}
\end{equation}
where $\eta_{p,a}$ is the net ring throughput, $\eta_{{\rm cor},a}$ is the coronal throughput into the same extraction statistic, $q_{\star,ka}$ and $q_{{\rm zodi},ka}$ are host-star and exozodiacal leakage terms, $r_{ka}$ is read noise, and $s_{ka}$ collects residual structured systematics.  After subtracting the modeled mean background, the Gaussian covariance approximation is
\begin{equation}
 C_{n,ka}\simeq t_{ka}\left[\eta_{p,a}Q_{{\rm exo},a}(\op_{ka}\truth)_k
 +\eta_{{\rm cor},a}Q_{{\rm cor},a}+q_{\star,ka}+q_{{\rm zodi},ka}+q_{{\rm dark},ka}\right]
 +N_{{\rm pix},ka}N_{{\rm fr},ka}\sigma_{r,a}^2 .
\label{eq:count_covariance}
\end{equation}
Multiplicative residuals on the bright coronal term add a calibration covariance,
\begin{equation}
 C_{{\rm flat},ij}\simeq \sigma_{\rm flat}^2 Q_{{\rm cor},i}Q_{{\rm cor},j}t_i t_j\,
 \rho_{\rm flat}(|\bm{x}_i-\bm{x}_j|),
\label{eq:flat_covariance}
\end{equation}
where $\rho_{\rm flat}$ is the spatial correlation function of the residual detector and extraction calibration error.  Coherent backgrounds may be represented as $\bkg=B\bm{\alpha}$, with template amplitudes $\bm{\alpha}$ and prior covariance $C_\alpha$, giving the effective covariance
\begin{equation}
 C_{\rm eff}=C_n+C_{\rm flat}+B C_\alpha B^\T .
\label{eq:effective_covariance}
\end{equation}
For a linear Gaussian static reconstruction with quadratic regularization $\Gamma$, the normal matrix and formal posterior covariance are
\begin{align}
 F = \op^\T C_{\rm eff}^{-1}\op+\Gamma, \qquad 
 C_{\truth|\data} \simeq F^{-1} .
\label{eq:posterior_covariance}
\end{align}
The current Fourier/Wiener calculation is the stationary, diagonal-covariance limit of this expression.  Eqs.~(\ref{eq:count_model})--(\ref{eq:posterior_covariance}) are retained because they are used directly below: Eq.~(\ref{eq:count_model}) defines the photon-count hierarchy summarized in Figure~\ref{fig:error_budget_control}; Eq.~(\ref{eq:flat_covariance}) motivates the calibration-leakage requirements in Sec.~\ref{sec:requirements}; and Eqs.~(\ref{eq:effective_covariance})--(\ref{eq:posterior_covariance}) explain why a mission-grade inversion must propagate explicit background and calibration covariance.  Future simulations should report not only image metrics but also the trace, diagonal, and selected eigenmodes of $C_{\truth|\data}$, because those modes are the statistically meaningful SGL error budget.

\begin{figure}[t]
\centering
\includegraphics[width=0.82\textwidth]{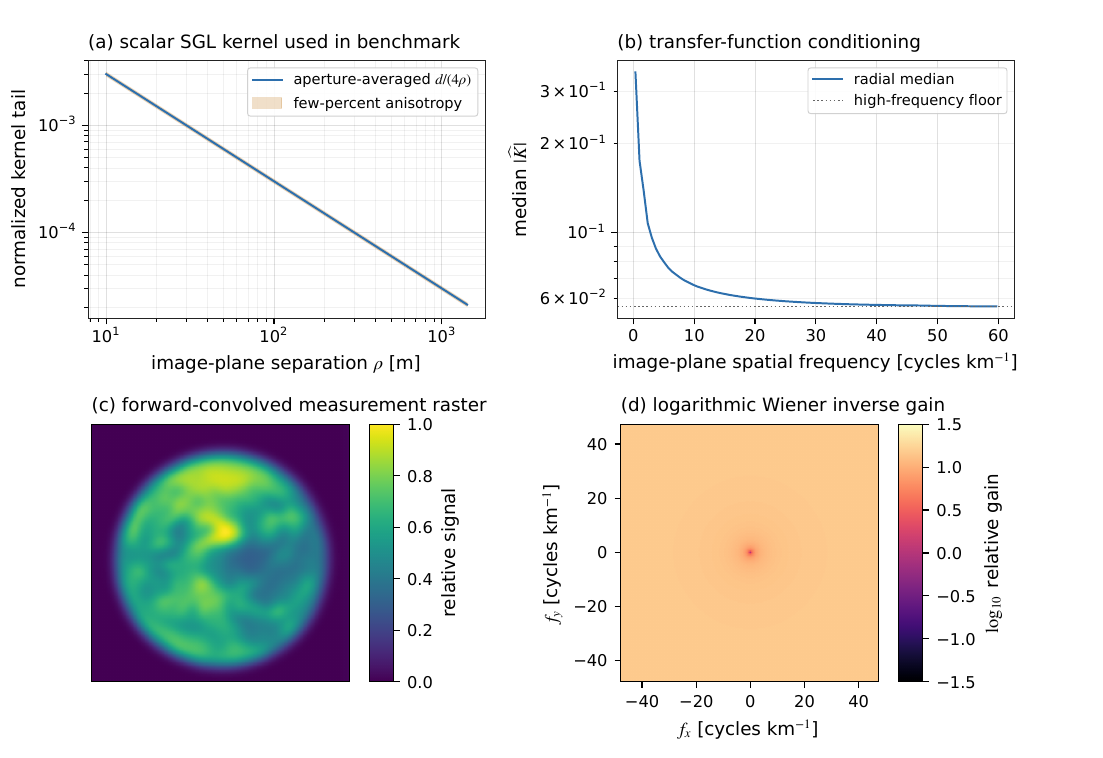}
\caption{SGL optical response used in the scalar simulation.  Panel (a) shows the aperture-averaged $d/(4\rho)$ kernel tail and the few-percent anisotropic envelope used for the multipole-mismatch sensitivity test.  Panel (b) shows the radial median of the discrete optical transfer function on the $192\times192$ guard-padded grid.  Panel (c) is the static SGL-convolved measurement raster generated from the real-Earth luminance image.  Panel (d) is the logarithmic Wiener inverse-filter gain for the mitigated scalar, phase-registered benchmark. The anisotropic envelope in panel~(a) is a controlled PSF-mismatch surrogate and is not computed from solar $J_\ell$ moments or from a physical multipole diffraction calculation.  Panels (a), (b), and (d) are model diagnostics; panel (c) is a forward-convolution output.}
\label{fig:optics}
\end{figure}

\begin{figure}[t]
\centering
\includegraphics[width=0.82\textwidth]{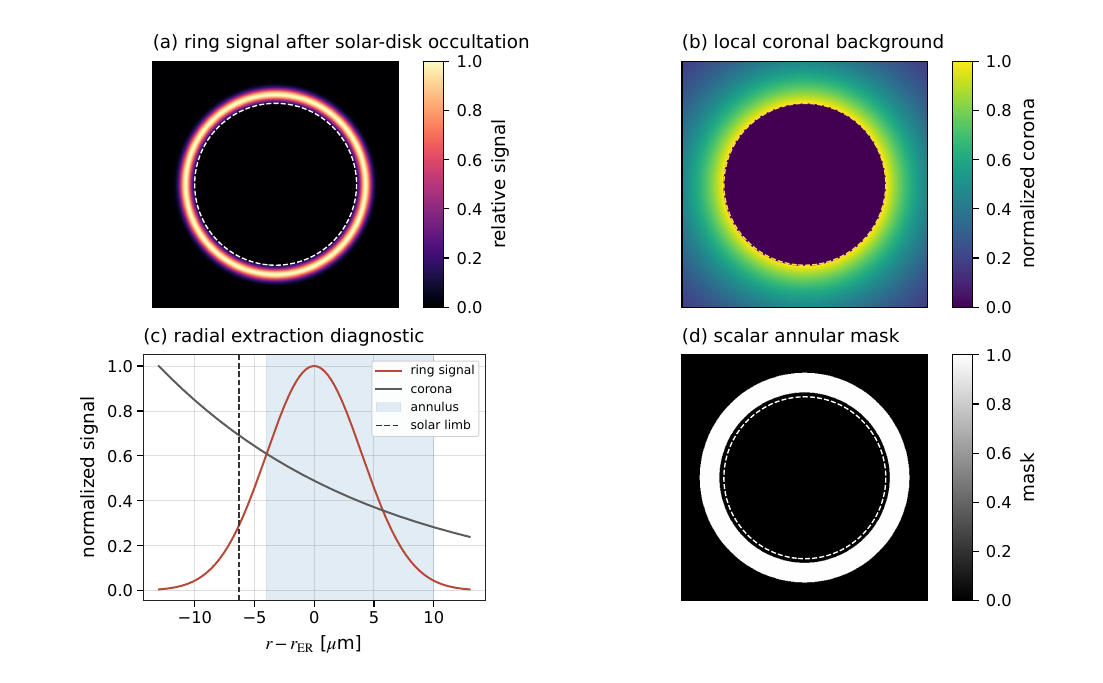}
\caption{Instrument-plane annular photometry diagnostic for
$z=650\,\AU$, $d=1\,\m$, $f=10\,\m$, and $\lambda=1\,\um$.
Panel (a) shows a diffraction-width Einstein-ring signal model after
occulting the solar disk.  Panel (b) shows the normalized coronal
background in the same focal-plane coordinates.  Panel (c) shows a radial diagnostic: the red curve is the ring profile, the gray curve is the local coronal background, the shaded region is the portion of the annular extraction aperture that remains outside the solar-limb mask plus guard margin, and the dashed line marks the solar limb relative to the Einstein-ring radius.  Panel (d) shows the corresponding annular mask used for aperture-integrated photometry.  This figure is an instrument-plane radiometric diagnostic, not a reconstructed planet image, and it does not replace a coronagraph or external-occulter propagation calculation.}
\label{fig:ring}
\end{figure}

\subsection{Fiducial observing case}

Table~\ref{tab:params} lists the fiducial parameters.  The $30\,\pc$ case is intentionally demanding; it is useful because the projected image is only $1.338\,\km$ across and therefore exposes the sensitivity to pixel spacing, detector calibration, temporal smearing, and navigation errors.

\begin{table}[t]
\centering
\caption{Fiducial numerical and observing parameters.}
\label{tab:params}
\begin{tabular}{ll}
\toprule
Quantity & Value \\
\midrule
Target distance $z_0$ & $30.0\,\pc$ \\
SGL observing distance $z$ & $650.0\,\AU$ \\
Telescope aperture $d$ & $1.0\,\m$ \\
Representative wavelength & $1.0\,\um$ \\
Planet reconstruction grid & $128\times128$ pixels \\
Computational grid & $192\times192$ pixels \\
Projected SGL image diameter & $1.338\,\km$ \\
Image-plane pitch & $10.46\,\m$ \\
Surface scale & $99.55\,\km\,\mathrm{pixel}^{-1}$ \\
Dwell per sample & $1800\,\s$ \\
Single-spacecraft cumulative dwell & $341.3\,\mathrm{d}$ \\
Exoplanet photon rate $Q_{\rm exo}$ & $8.01\times10^4\,\mathrm{s}^{-1}$ \\
Solar-corona photon rate $Q_{\rm cor}$ & $6.20\times10^9\,\mathrm{s}^{-1}$ \\
$Q_{\rm cor}/Q_{\rm exo}$ & $7.74\times10^4$ \\
Computed $\SNRc$ & $43.16$ \\
Spacing-penalty estimate & $0.0728$ \\
1800 s equatorial smear & $8.40$ source pixels \\
60 s equatorial smear & $0.28$ source pixels \\
\bottomrule
\end{tabular}
\end{table}

\section{Forward-model components}
\label{sec:forward}

\subsection{Photon rates and solar-corona shot noise}

The adopted broadband aperture-integrated rates for the fiducial optical case are \cite{TuryshevToth2022Resolved}
\begin{align}
Q_{\rm exo} &= 8.01\times10^4
\left(\frac{d}{1\,\m}\right)^2
\left(\frac{650\,\AU}{z}\right)^{1/2}
\left(\frac{30\,\pc}{z_0}\right)
\left(\frac{\lambda}{1\,\um}\right)\,\mathrm{s}^{-1},
\label{eq:Qexo}\\
Q_{\rm cor} &= 6.20\times10^9
\left(\frac{d}{1\,\m}\right)^2
\left(\frac{650\,\AU}{z}\right)^2
\left(\frac{\lambda}{1\,\um}\right)\,\mathrm{s}^{-1}.
\label{eq:Qcor}
\end{align}
The $Q_{\rm exo}$ normalization is the fiducial reflected-light Earth-analog rate from Ref.~\cite{TuryshevToth2022Resolved}: an Earth-radius planet at the adopted phase and Earth-like disk-averaged albedo observed in a representative optical/near-IR band centered at $\lambda=1\,\um$, with the effective aperture-integrated ring throughput absorbed into the scalar rate normalization.  In the count model of Eq.~\eqref{eq:count_model} this benchmark sets $\eta_p=\eta_{\rm cor}=1$; different albedo, phase, bandpass, or ring-extraction throughput values rescale $Q_{\rm exo}$ and $Q_{\rm cor}$ multiplicatively.

Figure~\ref{fig:optics} shows the implemented aperture-averaged kernel, the corresponding transfer-function conditioning, the static SGL-convolved measurement raster, and the regularized inverse-filter gain used by the scalar benchmark.

The convolved-image signal-to-noise ratio after dwell time $t$ is
\begin{equation}
\SNRc=\frac{Q_{\rm exo}t}{\sqrt{(Q_{\rm exo}+Q_{\rm cor})t}}.
\label{eq:snrc}
\end{equation}
For the fiducial parameters and $t=1800\,\s$, Eq.~\eqref{eq:snrc} gives $\SNRc=43.16$.  Since $Q_{\rm cor}t\simeq1.1\times10^{13}$ photons per sample, the Gaussian approximation to Poisson noise is excellent.  The mean corona is assumed to be estimated and subtracted; its shot noise remains and is the dominant stochastic term.

Figure~\ref{fig:ring} illustrates the instrument-plane annular photometry diagnostic used to motivate the scalar extraction statistic; it is a radiometric diagnostic, not a substitute for a coronagraphic or external-occulter propagation calculation.

The corresponding count-level SNR and error budget can be written in the same form used for conventional direct-imaging exposure calculations, but with different signal and background terms.  For one image-plane sample after subtracting the modeled mean background,
\begin{align}
 S_k &= t_k\eta_{p}Q_{\rm exo}(\op_k\truth)_k, \\
 B_k &= t_k\left(\eta_{\rm cor}Q_{\rm cor}+q_{\star,k}+q_{{\rm zodi},k}+q_{{\rm dark},k}\right), \\
 \SNR_k &= \frac{S_k}{\Big[S_k+B_k+N_{{\rm pix},k}N_{{\rm fr},k}\sigma_r^2+\sigma_{{\rm sys},k}^2\Big]^{1/2}} .
\label{eq:sample_snr_budget}
\end{align}
Here $\sigma_{{\rm sys},k}^2$ includes residual flat-field, nonlinearity, structured-background, and template-marginalization covariance.  Eq.~\eqref{eq:snrc} is the scalar fiducial limit of Eq.~\eqref{eq:sample_snr_budget} with $\eta_p=\eta_{\rm cor}=1$, negligible host/exozodi/detector rates, and $\sigma_{\rm sys}=0$.  In the corona-dominated limit the required dwell is
\begin{equation}
 t\simeq \SNRc^2\frac{Q_{\rm cor}}{Q_{\rm exo}^2},
\label{eq:corona_dominated_dwell}
\end{equation}
so the SGL trades the stellar-leakage and diffraction terms of conventional imaging for a bright, local solar-corona background that can be modeled in mean but not in shot noise.  This is the central SNR distinction between SGL imaging and remote coronagraphic or starshade imaging.  All scalar SNR values quoted below use the normalization $\eta_p=\eta_{\rm cor}=1$ in Eq.~\eqref{eq:count_model}; physical ring-extraction throughput, leakage covariance, and template marginalization enter through Eqs.~\eqref{eq:count_model}--\eqref{eq:effective_covariance} and must be demonstrated by a mission-level optical-extraction model.

Early critical assessments emphasized that SGL imaging is limited not only by reaching the focal region, but also by pointing, focal blur, solar-corona signal-to-noise, and the interpretation of the highly magnified image-plane geometry \cite{Landis2017,Willems2018}.  These concerns motivate the explicit treatment of coronal photon noise, image-plane sampling, and deconvolution penalty used below.

\subsection{Detector model}

The detector branch assumes that the annular Einstein-ring flux is measured over $8192$ detector pixels using short $0.02\,\s$ frames.  The per-frame corona level is $1.51\times10^4$ electrons per detector pixel, below the adopted $10^5$ electron full-well.  The simulation adds read noise, dark current, flat-field residuals, and residual quadratic nonlinearity.  Flat-field residuals multiply the bright corona and are therefore amplified in planet units by
\begin{equation}
\delta_{\rm planet}=\delta_{\rm flat}\frac{Q_{\rm cor}}{Q_{\rm exo}}.
\end{equation}
With $Q_{\rm cor}/Q_{\rm exo}=7.74\times10^4$, a $10\,\mathrm{ppm}$ effective residual corresponds to $0.774$ planet units and fails; a $0.1\,\mathrm{ppm}$ residual corresponds to $7.7\times10^{-3}$ planet units and is compatible with the photon-noise-limited branch.  This is one of the most stringent requirements revealed by the simulation.  Equivalently, a $1\,\mathrm{ppm}$ residual on the coronal signal corresponds to $7.7\times10^{-2}$ planet units in the fiducial case, so the relevant requirement is the end-to-end residual after annular photometry, coronal monitoring, detector calibration, and background-template subtraction, not the raw stability of any single detector parameter.

\subsection{Structured backgrounds and host-star leakage}

Structured coronal-subtraction residuals are represented by low- and mid-spatial-frequency fields normalized to a specified RMS in units of the mean convolved planetary signal.  Host-star leakage and exozodiacal light are represented by two arclet-like lobes plus a low-order gradient.  This nuisance-template model represents the coherent spatial structure that a physical leakage calculation would produce from host magnitude, planetary orbit, stellar leakage, and coronagraph/starshade geometry.  It is useful as an inverse-problem sensitivity test because it places coherent residual structure at spatial scales capable of producing false planetary features.

A useful scale check is that the host star's own SGL image cylinder is far from the planet's image cylinder for a normally separated planet.  For projected orbital separation $a_\perp$,
\begin{equation}
 \rho_{\star,{\rm img}}\simeq z\frac{a_\perp}{z_0}
 =1.57\times10^4
 \left(\frac{z}{650\,\AU}\right)
 \left(\frac{a_\perp}{1\,\AU}\right)
 \left(\frac{30\,\pc}{z_0}\right)\,\km .
\label{eq:host_star_offset}
\end{equation}
This is many orders of magnitude larger than the $\sim1\,\km$ planet image cylinder in the fiducial case.  Host-star contamination in the planet raster is therefore primarily an instrumental leakage, scattering, diffraction, or ring-extraction problem, not a coincident lensed image of the host star.  A physical leakage model should compute $q_{\star,k}$ in Eq.~\eqref{eq:count_model} from the stellar flux, off-axis angle, occulter/coronagraph propagation, detector scattering, and annular-sector mask, rather than from a normalized image-plane template alone.

\subsection{Finite exposure and planetary dynamics}

For an Earth-like rotation period $P_{\rm rot}$, the equatorial smear during one exposure is
\begin{equation}
 L_{\rm pix}(t_{\rm exp})=
 \frac{(2\pi R_\oplus/P_{\rm rot})t_{\rm exp}}{2R_\oplus/n}.
\label{eq:smear}
\end{equation}
At $n=128$, an $1800\,\s$ exposure produces $8.40$ source pixels of equatorial smear, or $836\,\km$.  A $60\,\s$ subexposure produces $0.28$ source pixels.  The simulation implements row-dependent east--west smear scaled by projected latitude.  A separate dynamic-raster sensitivity test assigns different rotational phases to different raster rows and advects the extracted cloud-like proxy.  This test intentionally violates the static convolution assumption and demonstrates that a single long raster is not an instantaneous image of a rotating cloudy planet.

The dynamic inverse problem can be stated explicitly.  Let $g$ denote the planet geometry and spin state, let $R(t;g)$ rotate surface coordinates into the observed frame, and decompose the radiance into persistent and transient terms,
\begin{equation}
 O(\lambda,t)=I(\lambda,t;g)\Big[R(t;g)O_{\rm surf}(\lambda)+O_{\rm cloud}(\lambda,t)\Big] ,
\label{eq:dynamic_scene}
\end{equation}
where $I$ contains illumination, phase, visibility, and limb weighting.  A minimal stochastic cloud model is
\begin{equation}
O_{\rm cloud}(t+\Delta t)=\Phi_c(\Delta t)O_{\rm cloud}(t)+w_c,
\qquad \langle w_cw_c^\T\rangle=Q_c(\Delta t),
\label{eq:cloud_state}
\end{equation}
with a correlation time and spatial covariance constrained by Earth observations or a general-circulation model.  The measurement at time $t_k$ is then
\begin{equation}
y_k=H_k(\eta)\,O(\lambda_k,t_k)+b_k(\eta)+n_k.
\label{eq:dynamic_measurement}
\end{equation}
The C3 failure case defined in Table~\ref{tab:metrics} is the result of applying a static inverse to data generated by Eq.~\eqref{eq:dynamic_measurement}.  It is a deliberately adverse estimator-mismatch test, not the expected observing mode of an SGL imaging campaign.

\subsection{Cloud mitigation and time-dependent reconstruction}
\label{sec:cloud_mitigation}

Clouds enter the SGL problem in two different ways.  For a persistent surface or albedo map, clouds are transient nuisance structure that must be separated from the rotating surface.  For atmospheric science, the same variations are a signal: cloud occurrence, cloud morphology, and temporal correlations are climate diagnostics.  The appropriate observing strategy therefore does not attempt to form one static image from one raster.  It acquires short, time-tagged subexposures, registers them to the known or jointly estimated spin phase, revisits each phase bin multiple times, and solves for persistent and transient components simultaneously.

A simple phase-registered multi-epoch model illustrates the mitigation.  Let $p$ index rotational phase bins and $e=1,\ldots,M_p$ index independent or partially independent visits to phase bin $p$.  After applying the known geometric registration operator $G_{pe}$, the data may be written schematically as
\begin{equation}
y_{pe}=H_{pe}\,G_{pe}O_{\rm surf}+H_{pe}O_{{\rm cloud},pe}+b_{pe}+n_{pe} .
\label{eq:phase_registered_data}
\end{equation}
A corresponding regularized estimator is
\begin{multline}
\min_{O_{\rm surf},\{O_{{\rm cloud},pe}\},\eta}
\sum_{p,e}\left\|C_{pe}^{-1/2}\left[y_{pe}-H_{pe}(\eta)G_{pe}O_{\rm surf}
        -H_{pe}(\eta)O_{{\rm cloud},pe}-b_{pe}(\eta)\right]\right\|^2 \\
+\lambda_s\|L_sO_{\rm surf}\|^2
+\lambda_c\sum_{p,e}\|L_cO_{{\rm cloud},pe}\|^2
+\lambda_t\sum_{p,e}\|O_{{\rm cloud},p,e+1}-\Phi_c O_{{\rm cloud},pe}\|^2
+R_\eta(\eta).
\label{eq:dynamic_inverse}
\end{multline}
Eq.~\eqref{eq:dynamic_inverse} is intentionally more general than the scalar Fourier inverse used in the C0--C9 benchmark, and it should be read as a required next validation step rather than as a completed component of the present pipeline.  A stronger mission-level test must solve a spin-resolved, time-dependent estimator that jointly recovers a rotating persistent surface map, transient cloud or atmospheric nuisance fields, PSF parameters, background templates, detector calibration, pointing, metrology, and ephemeris parameters from the time-tagged SGL data stream.  The C3b and C3c cases below are controlled, phase-registered coadd approximations to this problem; they support the plausibility of cloud mitigation under scalar assumptions, but they do not replace the full dynamic inversion in Eq.~\eqref{eq:dynamic_inverse}.

Several lower-cost mitigation strategies are contained in Eq.~\eqref{eq:dynamic_inverse}.  Cloud masking or downweighting is obtained by increasing the local variance in $C_{pe}$, or equivalently by applying weights
\begin{equation}
w_{pe}(x)=\left[\sigma_{n,pe}^2(x)+\sigma_{c,pe}^2(x)+\epsilon\right]^{-1},
\label{eq:cloud_weights}
\end{equation}
where $\sigma_{c,pe}^2$ is large for pixels, ring sectors, wavelengths, or time samples likely to be cloud dominated.  Robust alternatives replace the quadratic residual in Eq.~\eqref{eq:dynamic_inverse} with a Huber or Student-$t$ loss so that rare bright clouds do not bias the persistent map.  Statistical averaging follows when cloud residuals are approximately zero mean after masking and phase registration.  If $\sigma_c$ is the residual cloud fluctuation per phase-registered visit, then the contribution to the persistent-map variance scales as
\begin{equation}
\sigma_{\rm cloud,mean}^2(p)\simeq \frac{\sigma_c^2(p)}{N_{{\rm eff},p}},
\qquad
N_{{\rm eff},p}=\frac{M_p}{1+2\sum_{\ell=1}^{M_p-1}(1-\ell/M_p)\rho_\ell},
\label{eq:cloud_average}
\end{equation}
where $\rho_\ell$ is the lag-$\ell$ correlation of the cloud residuals in that phase bin.  Independent cloud realizations give the familiar $M_p^{-1/2}$ reduction in residual amplitude; correlated cloud fields reduce the effective number of samples rather than invalidating the imaging concept.

The C3b and C3c branches in Table~\ref{tab:metrics} are compact numerical checks of this mitigation path with an explicit observing interpretation.  In these branches, one registered cloud epoch means one full-SNR realization of the phase-registered image-plane sample set: each image-plane sample accumulates $t_{\rm samp}=1800\,\s$, implemented as $N_{\rm sub}=30$ subexposures of $t_{\rm sub}=60\,\s$, so that the per-subexposure equatorial smear is $L_{\rm pix}=0.28$ while the registered sample retains $\SNRc=43.16$.  The adopted cloud-recovery demonstrations use $M_p=16$ registered visits for the same rotational phase bin; they are therefore not sixteen subdivisions of one fixed $1800\,\s$ dwell.

C3b uses the surface-like and cloud-like diagnostic proxies in Fig.~\ref{fig:input}, generates $M_p=16$ phase-registered visits with independently advected cloud-proxy realizations, thresholds cloud-dominated proxy pixels as high-variance samples, and combines the registered reconstructions with a robust median.  Scored against the surface-like proxy, this estimator gives $\SNRr=7.21$, $\SSIM=0.813$, NRMSE $=0.572$, contrast recovery $=0.885$, and $\FRC_{50}=220\,\km$.  The cloud downweighting in C3b is derived from the diagnostic proxy, so the branch tests the recovery mechanism rather than the availability of an operational cloud classifier in a broadband scalar channel.

C3c uses the same scalar SGL operator, $\SNRc=43.16$ per registered visit, phase registration, cloud downweighting, Fourier/Wiener reconstruction, and robust coaddition as C3b, but replaces the advected cloud-proxy morphology with stochastic cloud fields generated independently of the surface proxy and independently across epochs.  The adopted realization uses $M_p=16$, $f_{\rm cl}=0.25$, $\ell_C=11$ pixels, a two-pixel dilation of the high-cloud downweighting mask, $\gamma=3.16\times10^{-2}$, and a weighted-Huber robust coadd.  Scored against the surface-like proxy, C3c gives $\SNRr=9.71$, $\SSIM=0.899$, NRMSE $=0.501$, contrast recovery $=1.028$, negligible mean bias, and a grid-limited $\FRC_{50}=199\,\km$.

Thus the C3 branch should be read as a requirement on temporal sampling and dynamic inversion.  C3 demonstrates that a one-pass static raster is the wrong estimator for a cloudy rotating planet.  C3b and C3c demonstrate, at scalar benchmark level, that phase registration, cloud masking or downweighting, and robust multi-epoch coaddition can recover persistent surface-proxy information when repeated visits are available.  They do not constitute a full spin-resolved dynamic inversion, physical cloud validation, or optical penetration through opaque clouds; in reflected light, persistent surface information is recovered statistically from repeated low-cloud or cloud-masked samples.  Validation against wavelength-dependent, time-tagged Earth observations, a general-circulation model coupled to radiative transfer, or physical cloud fields remains the mission-level cloud-recovery test.

\subsection{Navigation and pointing}

Image-plane navigation errors are modeled as coordinate warps.  The unmitigated case uses $4.0\,\m$ smooth drift and $1.5\,\m$ random jitter.  The mitigated case uses $0.25\,\m$ residual drift and $0.10\,\m$ residual jitter and incorporates the residual jitter as an effective blur kernel in the inverse model.  These values quantify the scale at which image-plane metrology becomes important.  For a $10.46\,\m$ sample pitch, meter-level errors are a significant fraction of a pixel and therefore reduce reconstructed contrast and spatial fidelity.

\begin{figure}[t]
\centering
\includegraphics[width=0.82\textwidth]{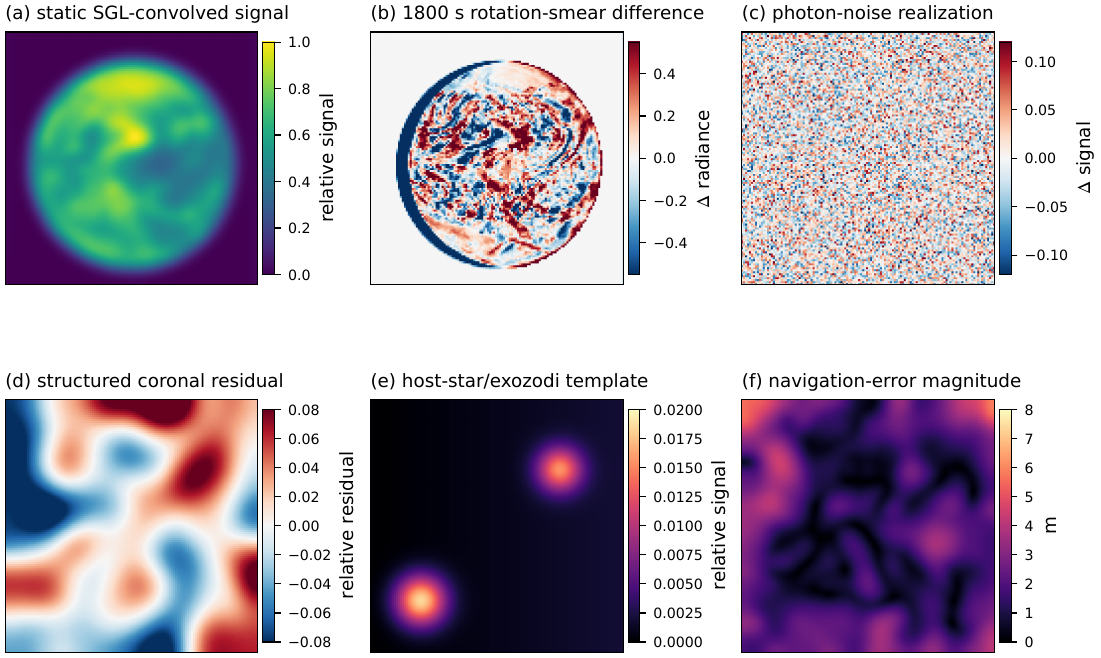}
\caption{Simulation-derived forward-model terms injected into the measurement.  Panel (a) is the static SGL-convolved signal.  Panel (b) is the source-radiance difference caused by one $1800\,\s$ rotational-smear dwell.  Panel (c) is a representative corona-dominated photon-noise realization.  Panel (d) is a structured coronal-subtraction residual.  Panel (e) is the host-star plus exozodiacal leakage template used as a coherent nuisance field.  Panel (f) is the unmitigated navigation-error magnitude in meters.  All panels are numerical arrays generated by the benchmark simulation and are displayed with their own color scales.}
\label{fig:terms}
\end{figure}

\subsection{Degradation mechanisms and mitigation methods}

Table~\ref{tab:effects} summarizes how each major effect enters the simulation and how it should be mitigated in a high-fidelity SGL imaging pipeline.  The key point is that most degradations are not independent scalar noise terms.  They either alter the effective optical operator, introduce coherent false structure, or destroy the stationarity assumption required for Fourier deconvolution.  The highest-quality reconstruction is obtained when the forward model is calibrated well enough that residuals are approximately stochastic, zero mean, and represented in the reconstruction covariance.

\begin{table*}[t]
\centering
\caption{Major image-quality degradations, controlling scales, implemented benchmark values, and mitigation requirements.  The table is organized by how each effect enters the measurement equation rather than by where it appears in a reconstructed image.  Surrogate entries are controlled sensitivity models rather than mission-grade physical models.}
\label{tab:effects}
\footnotesize
\renewcommand{\arraystretch}{1.18}
\setlength{\tabcolsep}{4pt}
\newcommand{\effa}[1]{\begin{minipage}[t]{0.11\textwidth}\raggedright #1\end{minipage}}
\newcommand{\effb}[1]{\begin{minipage}[t]{0.22\textwidth}\raggedright #1\end{minipage}}
\newcommand{\effc}[1]{\begin{minipage}[t]{0.30\textwidth}\raggedright #1\end{minipage}}
\newcommand{\effd}[1]{\begin{minipage}[t]{0.31\textwidth}\raggedright #1\end{minipage}}
\begin{tabular}{llll}
\toprule
\effa{Effect} & \effb{Controlling scale or equation} & \effc{Implemented benchmark result} & \effd{Mitigation and residual limitation} \\
\midrule
\effa{Monopole SGL blur} & \effb{Aperture-averaged kernel $K(0)=1$, $K(\rho>0)\simeq d/(4\rho)$; penalty $\SNRr/\SNRc\simeq0.891\Delta_{\rm img}/(dn)$.} & \effc{FFT convolution on a $192\times192$ guard grid. For $\Delta_{\rm img}=10.46\,\m$, $d=1\,\m$, $n=128$, the analytic penalty is $0.0728$.} & \effd{Calibrate the SGL+telescope ring-integral operator; upgrade to a full focal-plane diffraction model.} \\[0.7ex]
\effa{Solar multipoles} & \effb{Multipoles perturb the gravitational phase and produce anisotropic caustics; the imaging error is PSF mismatch.} & \effc{Energy-conserving $m=2,4$ sensitivity kernel with $\epsilon_2=0.03$, $\epsilon_4=0.01$, and $\theta_q=18^\circ$.} & 
\effd{Build and validate a wave-optical PSF library from measured or literature-based $J_2,J_4,\ldots$, solar spin orientation, impact parameter, wavelength, solar-plasma state, spacecraft ephemeris, and target ephemeris.  Current branch is a residual-sensitivity surrogate, not a physical multipole calculation.} \\[0.7ex]
\effa{Solar corona} & \effb{Mean can be modeled; shot noise remains. $\SNRc=Q_{\rm exo}t/[(Q_{\rm exo}+Q_{\rm cor})t]^{1/2}$.} & \effc{$Q_{\rm cor}/Q_{\rm exo}=7.74\times10^4$ and $\SNRc=43.16$ for $t=1800\,\s$. Corona photon noise alone gives $\SSIM=0.848$ and $\FRC_{50}=224\,\km$.} & \effd{Use annular masks, independent coronal monitoring, wavelength/polarization corona models, and residual covariance.} \\[0.7ex]
\effa{Host star/exozodi} & \effb{Coherent arclets and gradients create false low-frequency structure.} & \effc{Arclet peaks: $1.8\%$ of mean convolved planet signal before masking and $0.3\%$ after mitigation; exozodi gradients reduced from $0.6\%$ to $0.1\%$.} & \effd{Predict orbit-dependent leakage, mask contaminated sectors, and fit templates jointly. Current branch is a nuisance-template sensitivity model.} \\[0.7ex]
\effa{Detector response} & \effb{Multiplicative coronal residual: $\delta_{\rm planet}=\delta_{\rm flat}Q_{\rm cor}/Q_{\rm exo}$.} & \effc{$0.02\,\s$ frames over 8192 ring pixels give $1.51\times10^4$ corona electrons pixel$^{-1}$ frame$^{-1}$. $10$ ppm gives $0.774$ planet units; $0.1$ ppm gives $7.7\times10^{-3}$.} & \effd{Close an end-to-end residual budget after ring photometry, coronagraphic or external-occulter suppression, coronal monitoring, dark/flat/linearity calibration, template subtraction, detector covariance modeling, pointing reconstruction, and image-plane metrology.  The ppm values are system-level residual requirements, not isolated detector-stability specifications.}  \\[0.7ex]
\effa{Finite exposure} & \effb{Rotational smear $L_{\rm pix}=[(2\pi R_\oplus/P_{\rm rot})t_{\rm exp}]/(2R_\oplus/n)$.} & \effc{At $n=128$, $1800\,\s$ gives $8.40$ pixels ($836\,\km$) and $\FRC_{50}=607\,\km$; $60\,\s$ gives $0.28$ pixel.} & \effd{Use short time-stamped subexposures and phase-registered coadds.} \\[0.7ex]
\effa{Rotation/ clouds} & \effb{Long rasters sample different phases/cloud states and break stationarity.} & \effc{Static inversion of a deliberately time-incoherent raster gives $\SSIM=0.386$, NRMSE $=2.725$, and $\FRC_{50}=4247\,\km$.  The C3b advected-proxy recovery gives $\SSIM=0.813$ and $\FRC_{50}=220\,\km$ against the surface proxy; the C3c independent-cloud stress test gives $\SSIM=0.899$ and a grid-limited $\FRC_{50}=199\,\km$ against the same proxy.} & \effd{Use short subexposures, phase registration, cloud masking or downweighting, repeated/interleaved phase coverage, and a spin-resolved dynamic inversion separating persistent surface structure from transient clouds.  C3b/C3c are proxy/stochastic stress tests; physical validation requires time-tagged Earth observations or GCM+radiative-transfer scenes.} \\[0.7ex]
\effa{Navigation/\\ pointing} & \effb{Coordinate errors perturb the sampling operator; relevant scale is $\sigma_x/\Delta_{\rm img}$.} & \effc{With $\Delta_{\rm img}=10.46\,\m$, unmodeled $4.0\,\m$ drift and $1.5\,\m$ jitter degrade $\FRC_{50}$ to $386\,\km$.} & \effd{Use image-plane metrology, closed-loop scans, and measured coordinates in the forward operator.} \\[0.7ex]
\effa{Ephemeris/\\ geometry} & \effb{Center, scale, phase, spin-axis, period, and orbit errors create coherent coordinate and illumination errors.} & \effc{Not injected as an independent branch; represented indirectly through dynamic phases and navigation warps.} & \effd{Jointly estimate target geometry and the map using astrometry, repeated registration, and SGL revisits.} \\[0.7ex]
\effa{Regularization} & \effb{Small $|\widehat K|$ amplifies noise; large $\gamma$ oversmooths and biases contrast.} & \effc{Discrepancy-principle Fourier/Wiener inversion with support, positivity, and off-support subtraction.} & \effd{Replace with Bayesian/variational dynamic posterior over map, clouds, PSF, backgrounds, detector calibration, pointing, and ephemeris.} \\
\bottomrule
\end{tabular}
\end{table*}

The numerical forms of the principal injected forward-model terms are shown in Fig.~\ref{fig:terms}; these arrays are perturbations to the measurement model, not post-processing artifacts added to a reconstructed image.

\paragraph*{Simulation-case labels.} The case labels used below are defined quantitatively in Table~\ref{tab:metrics}.  They are introduced here to avoid ambiguity in the forward-model discussion: C0 is the static closure test; C1a--C1b isolate multipole-operator mismatch; C2a--C2b isolate exposure-smear mitigation; C3 is the deliberately time-incoherent rotating/cloudy raster; C3b is the phase-registered cloud-masked proxy-recovery demonstrator using advected diagnostic cloud-proxy epochs; C3c is the independent stochastic cloud-field stress test with cloud morphology generated independently of the surface proxy and independently across epochs; C4 is the corona photon-noise branch; C5a--C5b isolate detector/coronal calibration failure and mitigation; C6a--C6b isolate structured-background templates; C7a--C7b isolate navigation/control errors and mitigation; C8 is the mitigated scalar, phase-registered benchmark; and C9 applies the same scalar, phase-registered benchmark at higher photon statistics.  Only the visually most diagnostic cases are shown as reconstruction panels; all cases are included in the metric table and in Fig.~\ref{fig:metrics}.

\subsection{Error-source interpretation and residual limitations}

The individual error sources in Table~\ref{tab:effects} are best viewed as perturbations to the measurement equation, not as independent additions to an image after reconstruction.  Optical blur and jitter modify the operator $\op$; rotation, clouds, and ephemeris modify the time-dependent scene and geometry; the corona, host star, exozodi, and detector calibration enter as nuisance fields or noise covariance; and regularization controls how these uncertainties are amplified by the inverse.  This separation is important because the mitigation strategy must target the corresponding term in $\data=\op\truth+\bkg+\noise$, not only the final image.

Cloud cover and intrinsic time variability are the least complete physical components in the present simulation.  The cloud-like proxy is extracted from the real Earth image and advected in C3/C3b to test temporal incoherence and a minimal recovery path.  C3c adds one benchmark-level stress test by replacing the advected proxy morphology with independent stochastic cloud fields matched to the C3/C3b cloud proxy only in active-cloud covering fraction, amplitude distribution, and correlation length.  The result is nevertheless only a controlled scalar test: C3 violates the static convolution model and therefore produces a large reconstruction error, whereas C3b and C3c test whether phase registration, cloud downweighting, robust loss functions, and repeated phase coverage can recover persistent surface information under two different cloud-morphology surrogates.  These cases should not be interpreted as the expected performance of an SGL mission on cloudy planets.  They show that the estimator must match the data product.  For a persistent map, the appropriate model is a time-dependent inverse problem such as $O(x,t)=O_{\rm surf}(x_{\rm rot}(t))+O_{\rm cloud}(x,t)$, with priors on persistent surface structure, cloud correlation time, and cloud morphology.  Repeated phase coverage, multi-spacecraft interleaving, short subexposures, cloud masking or downweighting, robust loss functions, and state-space reconstruction are therefore the mechanism by which cloud variability is separated from map structure.  In our previous rotating-planet SGL work we reached the same qualitative conclusion: temporal dynamics must be represented in the forward model rather than treated as static image noise \cite{TothTuryshev2023Rotating}.  This conclusion is consistent with previous dynamic SGL imaging work, which showed that planetary rotation and orbital illumination must be represented in the forward model, and with the recent cloud-cover analysis \cite{Toth2025Clouds}, which highlights temporally varying clouds as a major practical requirement for high-fidelity SGL image recovery.

Finite exposure is better quantified.  Eq.~\eqref{eq:smear} shows that smear scales linearly with exposure time and inversely with desired linear image resolution.  At the fiducial $99.55\,\km$ pixel scale, an $1800\,\s$ dwell moves the equator by $836\,\km$, so the resulting image is necessarily blurred in longitude.  Splitting the dwell into $60\,\s$ frames reduces the displacement to $0.28$ pixel and permits phase registration.  The remaining limitation is not photon statistics, because the subframes can be coadded after registration, but accurate time stamping, spin-state knowledge, and cloud evolution during the dwell.

The solar corona sets the dominant stochastic background in the optical branch.  Eqs.~(\ref{eq:Qexo})--(\ref{eq:snrc}) implement the standard result that the mean corona can be estimated and removed, whereas its shot noise remains.  The large ratio $Q_{\rm cor}/Q_{\rm exo}=7.74\times10^4$ also turns small multiplicative calibration errors into significant false planetary signals.  The appropriate mitigation combines annular ring photometry, independent coronal monitoring, wavelength- or polarization-dependent corona models, and residual covariance in the inverse problem.  The spectrally resolved SGL analysis further shows why external occulters and longer wavelengths are important: they can relax internal-coronagraph diffraction constraints and reduce the relative impact of the corona in selected bands \cite{TuryshevToth2022Spectral}.

Host-star and exozodiacal contamination are modeled here as coherent nuisance templates because their main danger is false structure, not simply increased variance.  In a mission simulator, these terms should be computed from the host-star position relative to the planet's SGL optical axis, the planet orbit, the occulter or coronagraph response, and the ring-sector mask.  Mitigation requires predicting, masking, or jointly fitting the contaminated sectors of the Einstein ring.  The current template model is therefore adequate as an inverse-problem sensitivity test but not as an astrophysical leakage prediction.

Detector effects are dominated by calibration against the bright coronal background.  In the adopted photon-transfer model, dark current and read noise are subdominant, while multiplicative residuals obey $\delta_{\rm planet}=\delta_{\rm flat}Q_{\rm cor}/Q_{\rm exo}$.  This equation makes the requirement transparent: $10\,\mathrm{ppm}$ produces a false signal of order unity in planet units and fails, whereas $0.1\,\mathrm{ppm}$ is compatible with the mitigated simulation.  Saturation is controlled by short frames: the modeled corona level is $1.51\times10^4$ electrons per detector pixel per frame, giving a full-well margin of $6.6$ for a $10^5$ electron well.  Remaining detector work requires a covariance model for darks, flats, nonlinearity, bad pixels, cosmic rays, persistence, and interpixel coupling.

Navigation and pointing enter as errors in the sampling coordinates.  For the fiducial image-plane pitch, $4\,\m$ drift is $0.38$ pixel and $1.5\,\m$ jitter is $0.14$ pixel.  Such errors cannot be removed by image sharpening after the fact; they must be included in the forward operator.  The mitigated branch assumes residuals of $0.25\,\m$ and $0.10\,\m$, corresponding to $0.024$ and $0.010$ pixel, and includes residual jitter in the effective kernel.  The required tools are optical navigation, cross-scan metrology, repeated fiducial revisits, and an inversion that uses the measured coordinates rather than a nominal raster.

Solar quadrupole and higher-multipole perturbations are the principal optical-model limitation of the present work.  The surrogate kernel in Eq.~(\ref{eq:multipole_surrogate}) is useful because it tests the reconstruction sensitivity to anisotropic power in the SGL tail, but it is not a substitute for the wave-optical multipole calculations of an extended, rotating Sun.  The correct next step is a library of $K(\rho,\theta;\lambda,J_2,J_4,\ldots,\bm{s}_\odot,z)$ kernels and focal-plane ring responses.  Literature on the extended solar gravitational lens and multipole caustics provides the appropriate formal basis for this upgrade \cite{TuryshevToth2021ExtendedSun,TuryshevToth2023Faint}.

Regularization bias is unavoidable because deconvolution suppresses the signal components that are poorly conditioned by $\widehat K$.  The discrepancy-principle Wiener rule used here is objective within the simulation because it selects $\gamma$ from the expected noise level rather than the hidden truth.  It still biases high-frequency contrast when the data are noisy.  The validation criteria therefore include not only $\SNRr$ but also NRMSE, contrast recovery, photometric bias, structural similarity, and $\FRC_{50}$.  The Fourier-quotient and deconvolution-penalty literature justifies this baseline and shows why widely separated image-plane samples reduce the penalty relative to adjacent aperture-scale sampling \cite{TothTuryshev2021Recovery,TuryshevToth2022Resolved}.

Ephemeris and target-geometry errors remain the least explicitly modeled component after clouds.  Errors in the target orbit, planet center, projected radius, phase angle, spin axis, or rotation period create coherent coordinate and illumination errors.  To first order, a center error behaves like a translation of the image-plane raster; a scale error changes $D_{\rm img}$ and therefore $\Delta_{\rm img}$; a spin-axis or period error corrupts the phase registration used to control clouds and rotation.  The present paper recognizes these errors in $\op(\bm{\eta})$ but does not yet inject them as separate cases.  They should be included in the next-generation dynamic inverse model and constrained by pre-SGL astrometry, repeated SGL registration, and joint estimation of geometry with the map.

\section{Reconstruction and regularization}
\label{sec:reconstruction}

For stationary cases, reconstruction uses a Fourier/Wiener inverse,
\begin{equation}
\widehat{O}(\bm{f})=\frac{\widehat{K}^*(\bm{f})}{|\widehat{K}(\bm{f})|^2+\gamma}\widehat{y}(\bm{f}),
\label{eq:wiener}
\end{equation}
followed by off-support background subtraction, support projection, non-negativity, and conservative upper clipping.  The parameter $\gamma$ is selected without using the hidden truth.  Candidate reconstructions are reconvolved, and the selected solution is the one whose robust reconvolved residual is closest to the expected measurement noise, with weak penalties on excessive roughness and off-support leakage.  This discrepancy-principle rule avoids hand-tuning the inverse to maximize truth-known image quality.  The support mask is a geometric prior on the illuminated disk, not a prior on the surface albedo pattern.  In this benchmark the support is fixed by the preprocessed disk geometry, so all support-restricted metrics are conditional on correct center, radius, phase, and illumination support.  A flight analysis must estimate or marginalize these geometric parameters jointly with the map; the present support-constrained results should therefore not be interpreted as prior-free image-quality metrics.

The expected sampling-dependent deconvolution penalty is used as a validation check \cite{TuryshevToth2020Extended,TuryshevToth2022Resolved}:
\begin{equation}
\frac{\SNRr}{\SNRc}\simeq 0.891\frac{\Delta_{\rm img}}{dn},
\label{eq:penalty}
\end{equation}
where $n$ is the linear pixel count.  With $\Delta_{\rm img}=10.46\,\m$, $d=1\,\m$, and $n=128$, Eq.~\eqref{eq:penalty} gives $\SNRr/\SNRc=0.0728$.  This estimate captures the scaling of noise amplification; image metrics may differ because the reconstruction also uses support, positivity, regularization, a nonuniform Earth radiance field, and additional systematics.

\begin{figure}[t]
\centering
\includegraphics[width=0.82\textwidth]{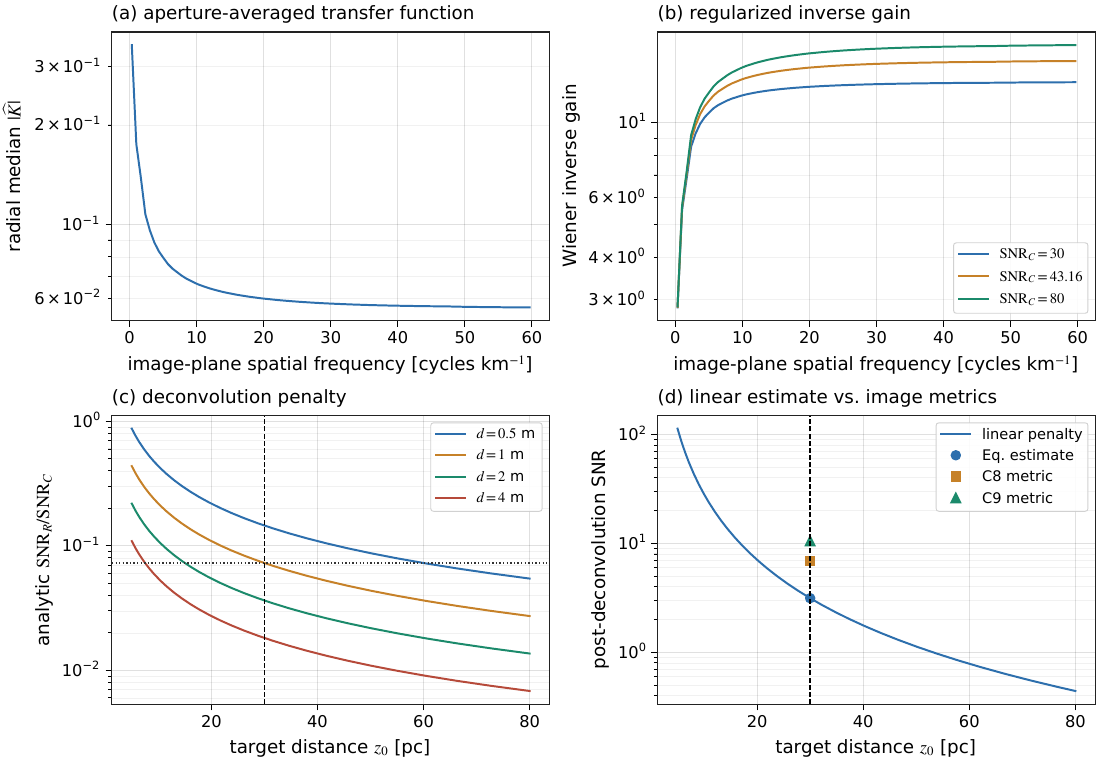}
\caption{Linear inverse-conditioning diagnostics computed from the adopted aperture-averaged kernel.  Panel (a) shows the radial median of the transfer function for the fiducial $192\times192$ guard-padded raster with $\Delta_{\rm img}=10.46\,\m$ and $d=1\,\m$.  Panel (b) shows how the Wiener inverse gain changes with the assumed convolved-image SNR.  Panel (c) evaluates the analytic deconvolution penalty $\SNRr/\SNRc$ from Eq.~(\ref{eq:penalty}) over target distance and aperture for an Earth-radius $n=128$ raster.  Panel (d) compares that linear estimate at $30\,\pc$ with the support-restricted C8 and C9 image metrics reported in Table~\ref{tab:metrics}.  The plot separates optical conditioning from the nonlinear support, positivity, and clipping operations used in the reported image metrics.}
\label{fig:inverse_conditioning}
\end{figure}

Figure~\ref{fig:inverse_conditioning} summarizes the transfer-function conditioning and separates the analytic deconvolution penalty from the support-restricted image metrics reported in Table~\ref{tab:metrics}.

For the fiducial parameters, Eq.~\eqref{eq:penalty} gives
\begin{equation}
 \left(\frac{\SNRr}{\SNRc}\right)_{\rm penalty}
 =0.891\frac{10.46}{1\times128}=0.0728,
 \qquad
 \SNRr\simeq3.14\quad(\SNRc=43.16).
\label{eq:penalty_fiducial}
\end{equation}
The Table~\ref{tab:metrics} image-domain values are larger because Eq.~\eqref{eq:penalty} is a linear noise-amplification estimate, whereas the reported $\SNRr$ is measured after support restriction, positivity, clipping, off-support background subtraction, and regularization.  These operations reduce the measured residual variance but also introduce bias and suppress poorly conditioned spatial modes.  The quantities should therefore be kept distinct: Eq.~\eqref{eq:penalty} gives the analytic sampling penalty; Eq.~\eqref{eq:posterior_covariance} defines the corresponding linear-estimator covariance diagnostic for a mission-level Gaussian error budget; and Eq.~\eqref{eq:snrr_metric} gives the support-restricted image-domain performance metric used in Table~\ref{tab:metrics}.  Consequently, the reported $\SNRr$ should be interpreted as a reconstruction-performance metric for the stated support-constrained estimator, not as a pure linear noise-equivalent SNR.

Eq.~\eqref{eq:wiener} is the transparent baseline estimator.  A mission-grade inverse extends it to the joint problem
\begin{equation}
\min_{\truth(t),\bm{\eta}}\left\|\bm{C}_{n}^{-1/2}\left[\data-\op(\bm{\eta})\truth(t)-\bkg(\bm{\eta})\right]\right\|^2+\mathcal{R}[\truth(t),\bm{\eta}],
\end{equation}
where $\bm{\eta}$ includes PSF multipoles, background templates, detector calibration, pointing, and ephemeris parameters.  This is a Bayesian or regularized dynamic inverse problem, not static deblurring.  In the cloud-mitigation limit described by Eq.~\eqref{eq:dynamic_inverse}, $O(t)$ is replaced by a persistent map plus a transient cloud state, and the reconstruction uses time tags, phase registration, masks or robust weights, and the measured scan coordinates rather than treating the raster as one simultaneous image.

\section{Validation metrics}
\label{sec:validation}

Metrics are computed on the planet support.  The post-reconstruction SNR is
\begin{equation}
\SNRr=\frac{\langle O\rangle}{\sigma(O_{\rm rec}-O)}.
\label{eq:snrr_metric}
\end{equation}
The normalized root-mean-square error is
\begin{equation}
\mathrm{NRMSE}=\frac{\sqrt{\langle(O_{\rm rec}-O)^2\rangle}}{\sigma(O)}.
\end{equation}
Contrast recovery is $\sigma(O_{\rm rec})/\sigma(O)$, and photometric bias is $(\langle O_{\rm rec}\rangle-\langle O\rangle)/\langle O\rangle$.  Structural similarity is computed using the disk image.  Spatial resolution is summarized by an FRC-like resolution proxy, $\FRC_{50}$, converted to kilometers using the $99.55\,\km$ pixel scale.  Stochastic branches are repeated in limited Monte Carlo ensembles to estimate random uncertainty; these ensembles do not include model-form uncertainty.

The $\FRC_{50}$ values reported here are resolution proxies rather than empirical half-data resolution measurements.  For a flight or end-to-end mission analysis, independent half-data reconstructions should be formed and $\FRC_{50}$ should be defined as the spatial scale at which their Fourier-ring correlation first falls below $0.5$.  In the present benchmark, deterministic closure cases and single-branch reconstructions are assigned a proxy crossing scale; values at $199\,\km$ are the two-pixel grid floor and should be interpreted only as a numerical resolution floor.  All $\FRC_{50}$ values are converted to physical scale using $2R_\oplus/n=99.55\,\km\,\mathrm{pixel}^{-1}$.

Validation tests include: (i) a noise-free convolution/deconvolution closure test; (ii) comparison with the analytic deconvolution penalty in Eq.~\eqref{eq:penalty}; (iii) stepwise controlled-degradation test of each effect; (iv) calibration-residual, exposure-smear, and photon-statistics sweeps; and (v) Monte Carlo reconstruction uncertainty maps.  The metric table includes two compact cloud-mitigation checks: C3b, which uses advected diagnostic cloud-proxy epochs, and C3c, which uses independent stochastic cloud fields matched to the proxy cloud statistics but independent of the persistent surface proxy and independent across epochs.  Both C3b and C3c are scored against the surface-like proxy rather than the full luminance truth.  They are benchmark-level stress tests of phase registration, masking/downweighting, and robust coaddition; full cloud-resolving dynamic inversion with physical cloud fields remains a mission-level extension.  These tests verify that the figures are simulation-based: the measurement rasters are convolved by the SGL operator, the noise and systematic arrays are injected into the data stream, and the reconstructions are produced by the stated inverse.

\section{Simulation results}
\label{sec:results}

\subsection{Stepwise controlled-degradation analysis}

\begin{figure}[p]
\centering
\includegraphics[width=0.88\textwidth]{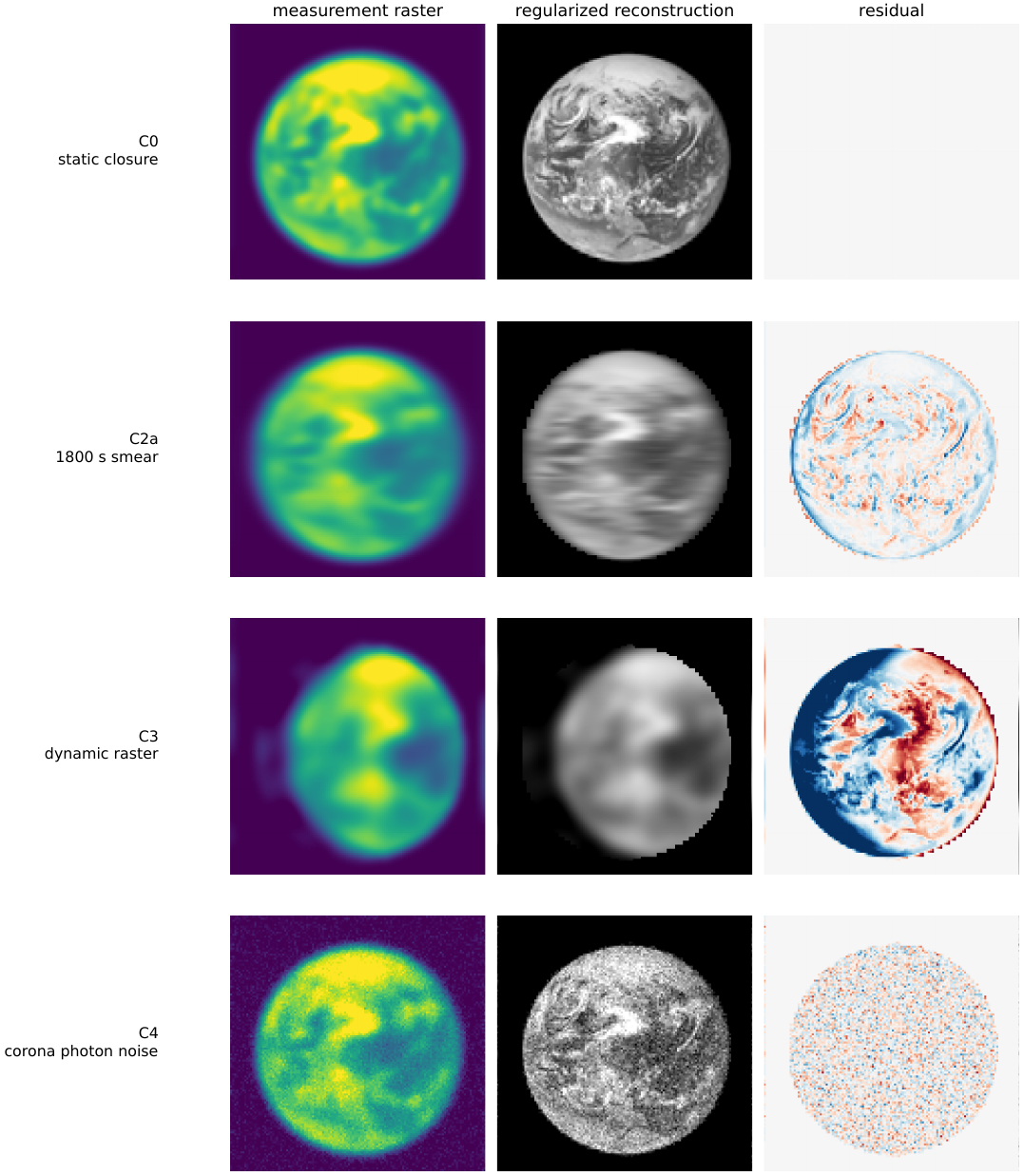}
\caption{Stepwise reconstructions for the static, finite-exposure, temporal-incoherence, and photon-noise branches.  Each row is a four-column block: case label, modeled image-plane measurement raster, regularized reconstruction, and residual relative to the same real-Earth truth.  C0 is the noise-free static closure test.  C2a adds a single $1800\,\s$ rotation-smear dwell.  C3 is a deliberately time-incoherent raster with changing rotational phase and an advected cloud-like proxy reconstructed with an intentionally inappropriate static inverse; it is a stress test of estimator mismatch, not a proposed SGL observing strategy.  C4 adds corona-dominated photon noise.}
\label{fig:reconA}
\end{figure}

\begin{figure}[p]
\centering
\includegraphics[width=0.88\textwidth]{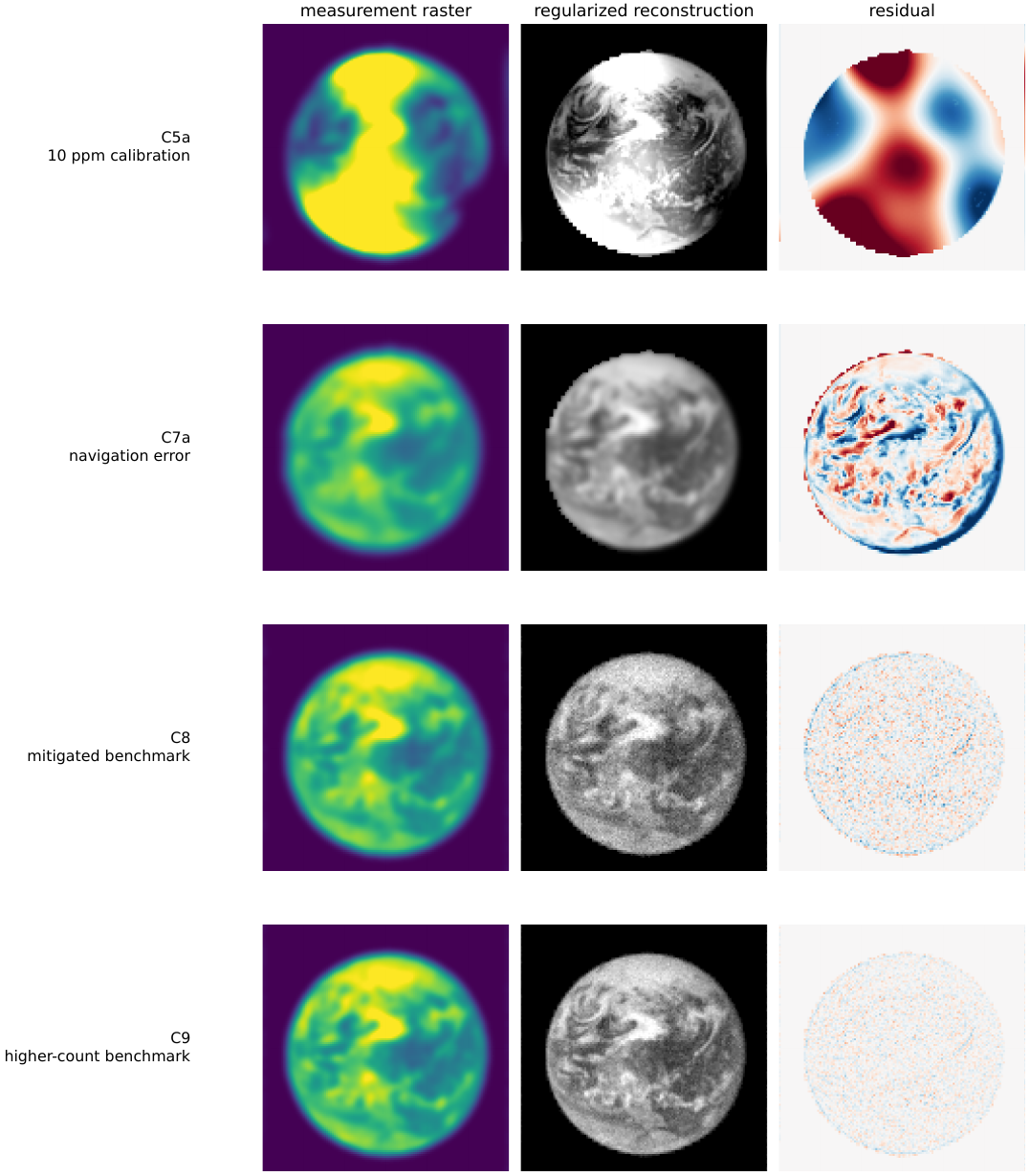}
\caption{Stepwise reconstructions for calibration failure, navigation error, and mitigation branches.  Each row is a four-column block: case label, modeled measurement raster, reconstruction, and residual relative to the same real-Earth truth.  C5a shows detector/corona calibration failure from a $10\,\mathrm{ppm}$ effective multiplicative residual on the bright corona.  C7a shows unmodeled navigation/control errors.  C8 is the mitigated scalar, phase-registered benchmark baseline at $30\,\pc$.  C9 increases photon statistics to $\SNRc=80$ while retaining the same mitigation strategy.}
\label{fig:reconB}
\end{figure}

\begin{figure}[t]
\centering
\includegraphics[width=0.82\linewidth]{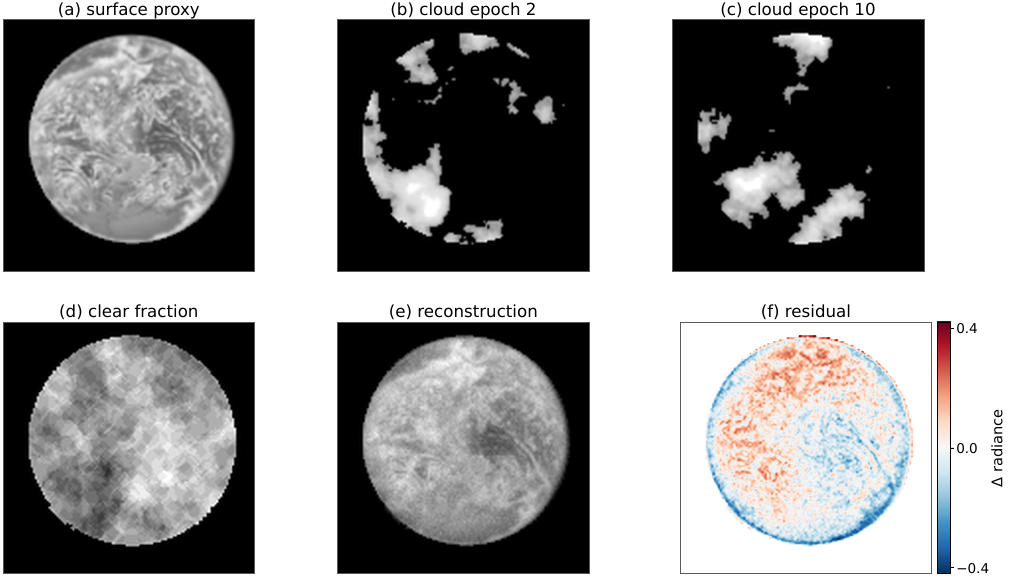}
\caption{Independent stochastic cloud-field stress test C3c.  Panel (a) is the surface-like proxy used as the persistent truth.  Panels (b) and (c) show two independently generated cloud epochs, produced independently of the surface proxy and independently across epochs while matching the C3/C3b cloud proxy in active-cloud covering fraction, brightness distribution, and correlation length.  Panel (d) shows the clear-sample fraction over the 16 registered epochs after cloud downweighting.  Panel (e) is the robust phase-registered reconstruction, and panel (f) is the residual relative to the surface-like proxy; the display scale is clipped at $\pm0.42$ in normalized radiance units. The branch uses $M_p=16$, $f_{\rm cl}=0.25$, $\ell_C=11$ pixels, and $\gamma=3.16\times10^{-2}$, and gives $\SNRr=9.71$, $\SSIM=0.899$, NRMSE $=0.501$, contrast recovery $=1.028$, bias $=0.0\%$, and a grid-limited $\FRC_{50}=199\,\km$.  C3c is an independent cloud-morphology stress test, not a physical cloud-radiative-transfer validation.}
\label{fig:c3c_independent_clouds}
\end{figure}

The stepwise results are summarized in Table~\ref{tab:metrics} and Figs.~\ref{fig:reconA}--\ref{fig:c3c_independent_clouds}.  The static calibrated case closes to numerical precision, as expected when the same forward and inverse operators are used.  The multipole sensitivity surrogate has little effect when calibrated and a small effect when unmodeled; this shows only that the adopted few-percent, energy-conserving anisotropic tail perturbation is not dominant in the scalar benchmark.  It does not validate the physical solar quadrupole, higher multipoles, or focal-plane caustic response, which remain separate mission-level PSF calibration inputs.

\begin{table}[t]
\centering
\caption{Stepwise reconstruction metrics.  Bias is reported in percent.  The column labeled $\SNRr$ is the support-restricted reconstruction metric of Eq.~\eqref{eq:snrr_metric}, not a pure linear noise-equivalent SNR.  The C0 SNR is omitted because the noise-free closure test produces a numerically singular residual standard deviation.  The C3b and C3c rows are controlled cloud-mitigation stress tests scored against the surface-like proxy, whereas the other rows are scored against the scalar luminance truth.}
\label{tab:metrics}
\setlength{\tabcolsep}{4pt}
\begin{tabular}{llrrrrrr}
\toprule
Case & Added realism / mitigation state & $\SNRr$ & SSIM & NRMSE & Contrast & Bias [\%] & $\FRC_{50}$ [km] \\
\midrule
C0 & static calibrated SGL & -- & 1.000 & 0.000 & 1.000 & -0.0 & 199 \\
C1a & multipole PSF unmodeled & 277.56 & 1.000 & 0.011 & 0.999 & 0.0 & 199 \\
C1b & multipole PSF calibrated & 517.35 & 1.000 & 0.006 & 0.996 & -0.1 & 199 \\
C2a & 1800 s finite-exposure smear & 6.01 & 0.782 & 0.487 & 0.855 & -1.4 & 607 \\
C2b & 60 s phase-registered subexposures & 92.34 & 0.999 & 0.032 & 0.987 & -0.1 & 199 \\
C3 & single raster temporal incoherence & 1.07 & 0.386 & 2.725 & 2.542 & 6.1 & 4247 \\
C3b$^{\dagger}$ & 16-epoch cloud-masked phase-registered coadd & 7.21 & 0.813 & 0.572 & 0.885 & 0.0 & 220 \\
C3c$^{\ddagger}$ & 16-epoch independent stochastic cloud-field coadd & 
9.71 & 0.899 & 0.501 & 1.028 & 0.0 & 199 \\
C4 & corona photon noise & 6.96 & 0.848 & 0.438 & 0.879 & -4.4 & 224 \\
C5a & detector/corona calibration failure & 1.32 & 0.483 & 2.228 & 2.196 & -9.8 & 4247 \\
C5b & detector calibrated & 6.74 & 0.843 & 0.449 & 0.868 & -4.1 & 232 \\
C6a & structured backgrounds unmitigated & 6.15 & 0.828 & 0.525 & 0.757 & -7.7 & 232 \\
C6b & structured backgrounds mitigated & 6.73 & 0.843 & 0.447 & 0.871 & -3.8 & 232 \\
C7a & navigation/control unmodeled & 5.34 & 0.764 & 0.585 & 0.761 & -7.2 & 386 \\
C7b & navigation/control mitigated & 6.82 & 0.846 & 0.442 & 0.864 & -3.8 & 232 \\
C8 & mitigated scalar, phase-registered benchmark, 30 pc & 6.89 & 0.848 & 0.439 & 0.890 & -4.0 & 232 \\
C9 & mitigated scalar, phase-registered high-count benchmark & 10.49 & 0.927 & 0.287 & 0.913 & -2.5 & 199 \\
\bottomrule
\end{tabular}
\begin{minipage}{1.00\linewidth}
\vspace{0.5ex}
\footnotesize Notes: $^{\dagger}$C3b isolates the cloud-recovery mechanism using the diagnostic surface-like proxy in Fig.~\ref{fig:input}; it is not a physical cloud-radiative-transfer simulation.  $^{\ddagger}$C3c uses independent stochastic cloud fields generated independently of the surface proxy and independently across epochs, with covering fraction, amplitude distribution, and correlation length matched to the C3/C3b cloud proxy; it is an independent cloud-morphology stress test, not physical cloud validation.  Values at $\FRC_{50}=199\,\km$ are the grid-limited two-pixel floor and should not be read as measured half-data resolution; larger values are finite FRC-like proxy crossing scales for the stated branch. The C3b and C3c metrics assume $M_p=16$ registered visits with the stated per-visit photon statistics, i.e., $\SNRc=43.16$ for $t_{\rm samp}=1800\,\s$ per image-plane sample in each visit.  If the same total $1800\,\s$ dwell were instead divided among 16 visits, the per-visit photon noise would increase by $\sqrt{16}$ and the C3b/C3c metrics would not apply.
\end{minipage}
\end{table}

Long-exposure smear is severe.  The $1800\,\s$ dwell reduces SSIM to $0.782$ and degrades $\FRC_{50}$ to $607\,\km$.  Phase-registered $60\,\s$ subexposures recover nearly the static result.  The single-raster temporal-incoherence branch gives the largest error: $\SSIM=0.386$, NRMSE $=2.725$, and $\FRC_{50}=4247\,\km$.  This is a model-identification result rather than a performance forecast.  It shows that a rotating cloudy exoplanet cannot be reconstructed as a static frame from a single long raster; it does not show that clouds are an unmitigated limitation for a multi-epoch, phase-registered SGL observing campaign.

C3b provides the corresponding minimal recovery check.  Using $16$ phase-registered cloud-proxy epochs, cloud-dominated pixel downweighting, and robust coaddition, it recovers the persistent surface-like proxy with $\SNRr=7.21$, $\SSIM=0.813$, NRMSE $=0.572$, contrast recovery $=0.885$, negligible mean bias, and $\FRC_{50}=220\,\km$.  In this compact demonstration the cloud downweighting is derived from the diagnostic proxy, so the branch tests the recovery mechanism rather than the availability of an operational cloud classifier in a broadband scalar channel.  Because C3b is scored against the surface-like proxy rather than the full luminance truth, it should be interpreted as a controlled demonstration of phase registration, masking/downweighting,  robust averaging, not as a mission-level cloud retrieval.

C3c gives the corresponding independent cloud-morphology stress test.  In this branch, 16 phase-registered epochs are generated with stochastic cloud fields that are independent of the surface proxy and independent across epochs, while matching the C3/C3b cloud proxy in active-cloud covering fraction, brightness distribution, and disk autocorrelation length.  The adopted realization uses $f_{\rm cl}=0.25$, $\ell_C=11$ pixels, $\gamma=3.16\times10^{-2}$, a two-pixel dilation of the high-cloud downweighting mask, and a weighted-Huber robust coadd.  Scored against the surface-like proxy, C3c gives $\SNRr=9.71$, $\SSIM=0.899$, NRMSE $=0.501$, contrast recovery $=1.028$, negligible mean bias, and a grid-limited $\FRC_{50}=199\,\km$.  The residual structure is dominated by the regularized scalar inverse, photon noise, and modest leakage from below-threshold cloud structure rather than by phase-registration failure.  Thus C3c reduces the specific benchmark-level vulnerability that C3b might be tuned to a single advected cloud proxy, while still remaining a synthetic cloud-morphology stress test rather than physical multi-epoch cloud validation.

Corona photon noise at $\SNRc=43.16$ produces a recognizable but noisy Earth image: $\SNRr=6.96$, $\SSIM=0.848$, and $\FRC_{50}=224\,\km$.  Detector/corona calibration failure is catastrophic: the $10\,\mathrm{ppm}$ effective flat-field branch reduces SSIM to $0.483$ and creates planet-scale false structure.  The calibrated detector branch with $0.1\,\mathrm{ppm}$ effective residual returns close to the photon-noise branch.

Navigation errors also matter.  The unmodeled $4\,\m$ drift plus $1.5\,\m$ jitter branch reduces SSIM to $0.764$ and $\FRC_{50}$ to $386\,\km$.  With metrology-informed residuals and a residual-jitter kernel, performance returns to the $\sim232\,\km$ branch.  The mitigated scalar, phase-registered benchmark (C8), which assumes calibrated scalar backgrounds, residual navigation errors included in the effective kernel, short-subexposure phase registration, and no unresolved one-pass cloud aliasing, gives $\SNRr=6.89$, $\SSIM=0.848$, NRMSE $=0.439$, contrast recovery $=0.890$, and photometric bias $=-4.0\%$.  The high-count branch (C9) applies the same scalar, phase-registered mitigation strategy and improves to $\SNRr=10.49$, $\SSIM=0.927$, NRMSE $=0.287$, and $\FRC_{50}=199\,\km$.

The remaining table entries are included because they isolate specific requirements even when their images are visually redundant.  C1a and C1b test whether the few-percent multipole surrogate is known or unmodeled; both remain near the numerical floor, so they are not shown as separate reconstruction rows.  C2b demonstrates that short, phase-registered subexposures largely remove the finite-exposure penalty visible in C2a.  C5b, C6a, C6b, and C7b separate the detector, structured-background, and navigation terms that are combined in C8: their similar SSIM and $\FRC_{50}$ values show that after mitigation these terms are subdominant to photon statistics and residual regularization bias.  Thus Figures~\ref{fig:reconA}--\ref{fig:reconB} show the most diagnostic image failures and the two benchmark recoveries, while Table~\ref{tab:metrics} and Fig.~\ref{fig:metrics} retain the full sensitivity bookkeeping.

\subsection{Metric trends and uncertainty}

\begin{figure}[t]
\centering
\includegraphics[width=0.86\linewidth]{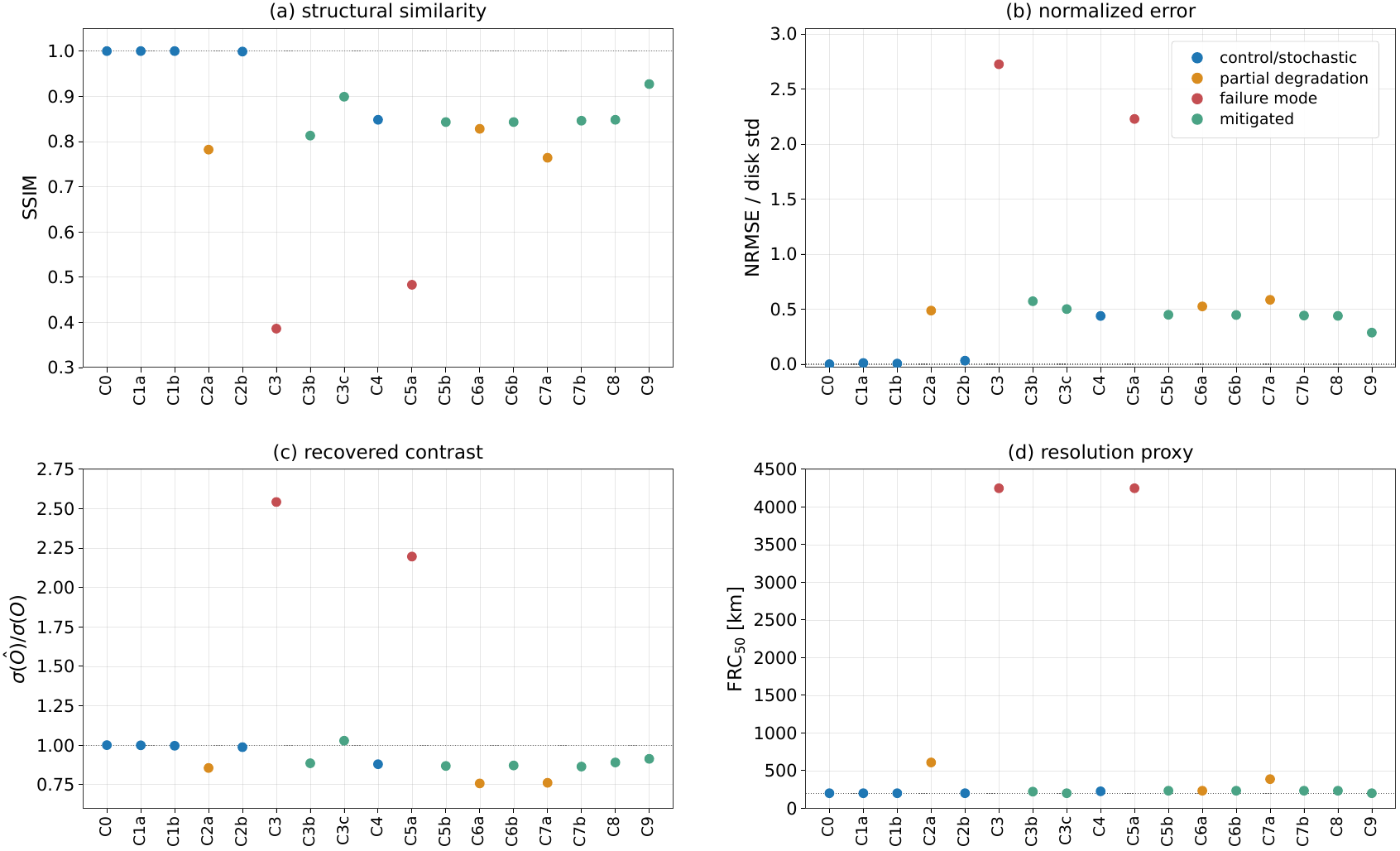}
\caption{Image-quality metrics for the discrete simulation cases.  The markers are independent cases in the same order as Table~\ref{tab:metrics}; no connecting lines are used because the cases are controlled simulation branches rather than samples of a continuous parameter.  The four panels show structural similarity, normalized error, recovered contrast, and the $\FRC_{50}$ resolution proxy.  The dashed lines mark the ideal $\SSIM$ and contrast values, zero normalized error, and the two-pixel grid-limited resolution floor.  The figure includes all cases in Table~\ref{tab:metrics}, including cases not shown as reconstruction panels.  The marker colors distinguish control/stochastic branches, partial degradations, deliberately mismatched or uncalibrated failure modes, and mitigated branches including the C3b advected-proxy recovery, the C3c independent-cloud-morphology stress test, and the C8--C9 scalar benchmarks.}
\label{fig:metrics}
\end{figure}

Figure~\ref{fig:metrics} gives $\SSIM$, NRMSE, contrast recovery, and $\FRC_{50}$ for all cases.  The metric pattern is more informative than any single scalar.  C3 and C5a are qualitatively different failure modes: C3 introduces temporal striping and cloud/rotation aliasing, while C5a introduces broad false photometric structure from calibration residuals.  C3b shows that phase registration, cloud downweighting, and robust multi-epoch coaddition recover persistent surface information in the advected-proxy test.  C3c tests the same mitigation path with cloud morphology generated independently of the surface proxy and independently across epochs.  Its recovery remains far from the C3 static-raster failure mode, reducing the possibility that the cloud-mitigation result is tuned to a single advected cloud proxy.  C7a is less catastrophic than C3 or C5a but degrades resolution and contrast.  C8 and C9 show that useful maps can be recovered within the scalar, phase-registered benchmark when temporal sampling, calibration, background suppression, and navigation are all mitigated.

\begin{figure}[t]
\centering
\includegraphics[width=0.86\textwidth]{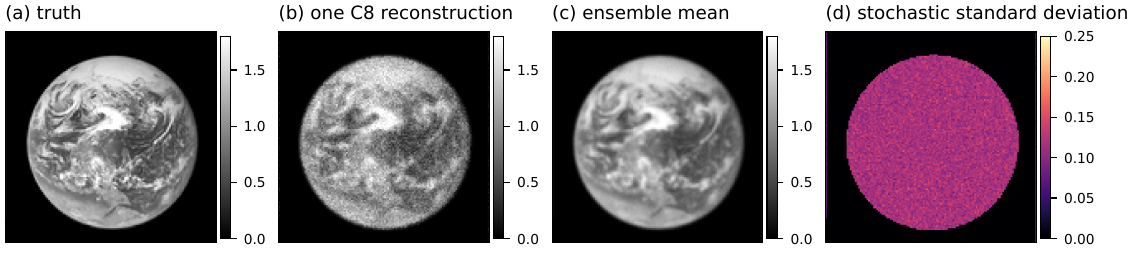}
\caption{Uncertainty propagation for the mitigated scalar, phase-registered C8 branch.  Panel (a) is the real-Earth truth.  Panel (b) is one C8 reconstruction from that branch.  Panel (c) is the mean of a 48-realization stochastic ensemble generated with the same scalar forward model.  Panel (d) is the per-pixel standard-deviation map under that stochastic model.  The ensemble quantifies random noise under fixed model assumptions; it does not include physical model uncertainty from clouds, multipoles, occulter propagation, or navigation dynamics.}
\label{fig:uncertainty}
\end{figure}

Uncertainty propagation is illustrated in Fig.~\ref{fig:uncertainty}.  The Monte Carlo standard-deviation map is largest where the reconstructed radiance field has high spatial gradients and where residual noise is amplified by the inverse filter.  This is stochastic uncertainty only; it does not include model-form uncertainty in the multipole PSF, coronal templates, detector calibration, or the target's true temporal evolution.

\section{Requirements and sensitivity}
\label{sec:requirements}

The simulation identifies five dominant requirements.

First, exposure time must be short compared with the time required for a surface feature to move by a source pixel.  For the fiducial grid, an $1800\,\s$ dwell smears equatorial features by $8.40$ pixels; a $60\,\s$ subexposure smears them by $0.28$ pixel.  High-quality imaging therefore requires time-tagged subexposures and phase registration.  A long dwell may be accumulated only after correcting for planetary rotation and scan geometry.

Second, clouds and intrinsic variability require redundant phase coverage rather than a one-pass static raster.  A persistent surface map is obtained by repeatedly observing the same rotational phase, registering each visit to the spin geometry, masking or downweighting cloud-dominated samples, and solving a dynamic inverse problem such as Eq.~\eqref{eq:dynamic_inverse}.  If the residual cloud amplitude after masking is $\sigma_c$ and the allowed contribution to the persistent-map uncertainty is $\epsilon_c\sigma_O$, then independent visits require approximately
\begin{equation}
 M_p \gtrsim \Big(\frac{\sigma_c}{\epsilon_c\sigma_O}\Big)^2,
\label{eq:cloud_epochs}
\end{equation}
with $M_p$ replaced by $N_{{\rm eff},p}$ from Eq.~\eqref{eq:cloud_average} when cloud realizations are correlated.  This requirement is an observing-design requirement, not a loss of SGL angular resolution.

Third, effective corona/detector calibration must be at the sub-ppm level after annular averaging and background modeling.  Since $Q_{\rm cor}/Q_{\rm exo}=7.74\times10^4$, ppm-level residuals are not small in planet units.  The requirement is not merely detector flat-field stability; it is the end-to-end residual after ring photometry, coronagraphic or external-occulter suppression, coronal monitoring, dark/flat calibration, linearity correction, detector covariance modeling, background-template subtraction, pointing reconstruction, and image-plane metrology.  The present benchmark identifies the numerical residual levels that the scalar reconstruction can tolerate; a mission-level demonstration must close a credible calibration and metrology chain showing that these residuals remain below the allocated covariance in Eqs.~\eqref{eq:count_model}--\eqref{eq:effective_covariance}.

Fourth, image-plane metrology must keep residual errors well below the $10.46\,\m$ sample pitch.  In the present model, meter-level unmodeled drift and jitter noticeably reduce resolution, while residuals of order $0.1$--$0.25\,\m$ are compatible with the mitigated scalar, phase-registered branch when included in the effective PSF.

Fifth, the inverse model must contain the correct optical kernel.  The present multipole surrogate is mild, but the literature shows that a realistic extended rotating Sun can form caustic structures that must be included in the inverse operator \cite{TuryshevToth2021ExtendedSun,TuryshevToth2023Faint}.  PSF calibration and solar-ephemeris modeling are therefore fundamental science requirements.

The main requirements can be written as explicit design inequalities.  If the allowed false additive signal from a multiplicative coronal residual is $\delta_{\rm max}$ in planet units, then
\begin{equation}
 \delta_{\rm flat}<\delta_{\rm max}\frac{Q_{\rm exo}}{Q_{\rm cor}}
 =0.129\left(\frac{\delta_{\rm max}}{0.01}\right)
 \left(\frac{30\,\pc}{z_0}\right)
 \left(\frac{z}{650\,\AU}\right)^{3/2}\,\mathrm{ppm}.
\label{eq:flat_req}
\end{equation}
If residual image-plane errors must be below a fraction $\epsilon_x$ of a sample, then
\begin{equation}
 \sigma_x<\epsilon_x\Delta_{\rm img}
 =0.314\left(\frac{\epsilon_x}{0.03}\right)
 \left(\frac{R_p}{R_\oplus}\right)
 \left(\frac{128}{n}\right)
 \left(\frac{z}{650\,\AU}\right)
 \left(\frac{30\,\pc}{z_0}\right)\,\m .
\label{eq:nav_req}
\end{equation}
The corresponding apparent sky-angle tolerance is
\begin{equation}
 \sigma_\theta < \frac{\sigma_x}{z}
 =0.66\left(\frac{\sigma_x}{0.31\,\m}\right)
 \left(\frac{650\,\AU}{z}\right)\,\nas,
\label{eq:sky_angle_req}
\end{equation}
which is too stringent to treat as a prelaunch catalog-astrometry requirement alone; it must be closed by in-situ image-plane acquisition, repeated registration, and measured-coordinate inversion.  Requiring equatorial smear below $\epsilon_L$ source pixels gives
\begin{equation}
 t_{\rm exp}<\frac{\epsilon_L P_{\rm rot}}{\pi n}
 =64.3\left(\frac{\epsilon_L}{0.30}\right)
 \left(\frac{P_{\rm rot}}{23.934\,\mathrm{h}}\right)
 \left(\frac{128}{n}\right)\,\s .
\label{eq:texp_req}
\end{equation}
These equations show why the apparently simple $1800\,\s$ dwell must be decomposed into short, time-tagged subexposures and why wall-clock observing time motivates parallel or interleaved sampling:
\begin{equation}
 N_{\rm sc}\gtrsim11.4
 \left(\frac{n}{128}\right)^2
 \left(\frac{t}{1800\,\s}\right)
 \left(\frac{30\,\mathrm{d}}{T_{\rm cal}}\right),
\label{eq:nsc_req}
\end{equation}
for a complete raster in calendar time $T_{\rm cal}$; in practice the scan should be interleaved across rotational phase bins so that the same calendar time also builds the redundancy required by Eq.~\eqref{eq:cloud_epochs}.

\begin{figure}[t!]
\centering
\includegraphics[width=0.82\textwidth]{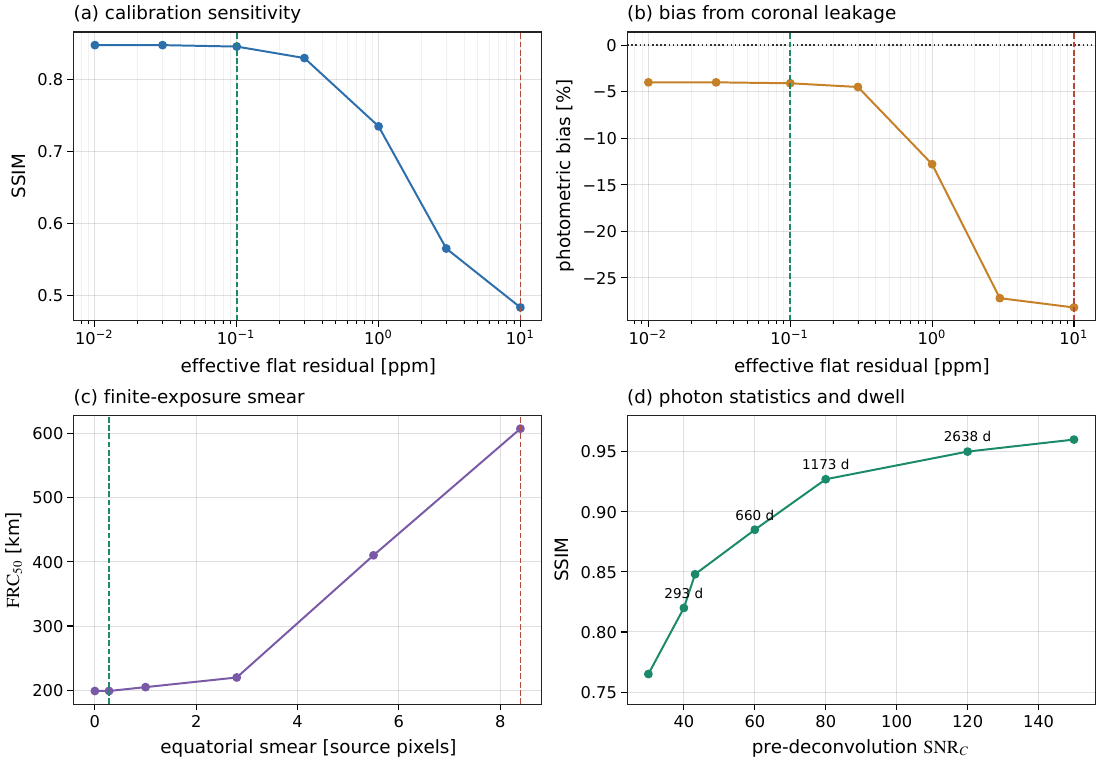}
\caption{Requirement sweeps generated from the numerical trends used in Table~\ref{tab:metrics}.  Panels (a) and (b) use the calibration-residual sequence where an effective multiplicative residual on the bright corona is converted to planet units through $\delta_{\rm planet}=\delta_{\rm flat}Q_{\rm cor}/Q_{\rm exo}$.  Panel (c) evaluates the finite-exposure sequence using the smear variable in Eq.~\eqref{eq:smear}; the vertical guides mark the 60 s and 1800 s cases.  Panel (d) uses the photon-statistics sequence, with representative single-spacecraft cumulative dwell times annotated.  The sweeps isolate scalar requirements; cloud mitigation additionally requires the repeated phase coverage described by Eqs.~\eqref{eq:dynamic_inverse} and \eqref{eq:cloud_epochs}.}
\label{fig:requirements}
\end{figure}

For a cloudy rotating planet the dwell requirement must include the number of registered visits per phase bin.  Let $M_p$ be the number of visits used to suppress cloud residuals in phase bin $p$, let $N_\phi$ be the number of rotational phase bins retained as separate map products, and let $f_{\rm oh}\ge1$ denote slew, calibration, communication, and scheduling overhead.  The photon-time part of the wall-clock observing budget is then
\begin{equation}
T_{\rm obs}\simeq
f_{\rm oh}\,\frac{N_\phi M_p n^2 t_{\rm samp}}{N_{\rm sc}}
=
341.3\,{\rm d}\,
f_{\rm oh}\,
\left(\frac{N_\phi M_p}{N_{\rm sc}}\right)
\left(\frac{n}{128}\right)^2
\left(\frac{t_{\rm samp}}{1800\,\s}\right) .
\label{eq:cloud_calendar_budget}
\end{equation}
Equivalently,
\begin{equation}
N_{\rm sc}\gtrsim
15.0\,f_{\rm oh}\,N_\phi
\left(\frac{M_p}{16}\right)
\left(\frac{n}{128}\right)^2
\left(\frac{t_{\rm samp}}{1800\,\s}\right)
\left(\frac{365\,{\rm d}}{T_{\rm cal}}\right),
\label{eq:cloud_nsc_budget}
\end{equation}
with the coefficient becoming $60.7$ for $T_{\rm cal}=90\,{\rm d}$ and $182$ for $T_{\rm cal}=30\,{\rm d}$.  The C3b/C3c demonstrations correspond to $M_p=16$ and $N_\phi=1$ for one registered broadband map.  Thus the cloud-recovery benchmark requires $16\times341.3\,{\rm d}=14.95\,{\rm yr}$ of single-spacecraft cumulative dwell per registered phase bin before overhead, or about $341\,{\rm d}$ with $N_{\rm sc}=16$, $171\,{\rm d}$ with $N_{\rm sc}=32$, and $85\,{\rm d}$ with $N_{\rm sc}=64$.

The cadence requirement is separate from the photon-time requirement.  The subexposure must satisfy Eq.~\eqref{eq:texp_req}; for the fiducial Earth-rotation case, $t_{\rm sub}=60\,\s$ keeps $L_{\rm pix}=0.28$.  Repeated visits to the same phase bin should be separated by an integer number of rotations,
\begin{equation}
\Delta t_{\rm revisit}=qP_{\rm rot},\qquad
q\ge \max\left[1,\left\lceil\frac{\tau_c}{P_{\rm rot}}\right\rceil\right],
\label{eq:cloud_revisit_cadence}
\end{equation}
when independent cloud realizations are desired, where $\tau_c$ is the relevant cloud-residual correlation time after masking or downweighting.  The minimum calendar span for $M_p$ statistically independent visits to one phase bin is therefore approximately $(M_p-1)\Delta t_{\rm revisit}$, but for the fiducial photon budget the raster dwell in Eq.~\eqref{eq:cloud_calendar_budget} is usually the controlling term unless many spacecraft are used.  For a binary clear/cloud approximation with independent cloud cover fraction $f_{\rm cl}$, the expected number of usable low-cloud samples per surface element is
\begin{equation}
N_{\rm clear}\simeq M_p(1-f_{\rm cl}) ,
\label{eq:clear_samples}
\end{equation}
before replacing $M_p$ by $N_{{\rm eff},p}$ to account for cloud correlations.  In the C3c stress test, $M_p=16$ and $f_{\rm cl}=0.25$ correspond to approximately $12$ clear or low-cloud samples per registered phase bin before correlation losses.

Table~\ref{tab:cloud_observing_budget} summarizes the observing-time interpretation of Eqs.~\eqref{eq:cloud_calendar_budget}--\eqref{eq:cloud_nsc_budget} for one registered scalar map, the C3b/C3c cloud-recovery branch, and phase-resolved cloud-aware products.

\begin{table}[h!]
\centering
\caption{Fiducial \(d=1\,{\rm m}\) observing-time interpretation for a \(128\times128\) Earth-radius map at \(30\,\pc\) with \(t_{\rm samp}=1800\,\s\) per image-plane sample. The values exclude overhead and assume the fiducial scalar ring-extraction throughput and photon-rate normalization used in Eqs.~\eqref{eq:Qexo}--\eqref{eq:Qcor}.  For phase-resolved products with $N_\phi>1$, multiply the dwell by $N_\phi$.}
\label{tab:cloud_observing_budget}
\begin{tabular}{lcccc}
\toprule
Product or benchmark & $M_p$ & single-spacecraft dwell & $T_{\rm obs}$ for $N_{\rm sc}=16$ & $N_{\rm sc}$ for $T_{\rm cal}=1\,{\rm yr}$ \\
\midrule
One registered scalar map & 1  & $341\,{\rm d}$        & $21.3\,{\rm d}$  & 1 \\
C3b/C3c cloud-recovery branch & 16 & $14.95\,{\rm yr}$ & $341\,{\rm d}$   & 15 \\
Phase-resolved cloud-aware product & $16$ & $14.95\,N_\phi\,{\rm yr}$ & $341\,N_\phi\,{\rm d}$ & $15N_\phi$ \\
\bottomrule
\end{tabular}
\end{table}

Figure~\ref{fig:requirements} shows the calibration, finite-exposure, and photon-statistics sweeps corresponding to the scalar requirements in Eqs.~\eqref{eq:flat_req}, \eqref{eq:texp_req}, and \eqref{eq:nsc_req}; these sweeps isolate scalar sensitivities, while the rotating-cloudy case additionally requires the phase-registered redundancy described by Eqs.~\eqref{eq:cloud_calendar_budget}--\eqref{eq:clear_samples}.

\section{Mission architectures and observing strategies}
\label{sec:architectures}

The SGL observing problem favors architectures that reduce temporal aliasing and accumulate redundant samples.  A single spacecraft scanning the fiducial $128\times128$ raster at $1800\,\s$ per sample requires $341\,\mathrm{d}$ of cumulative dwell, excluding slews, calibrations, communication scheduling, and overhead.  This is a long but concrete observing requirement, not an undefined optical limitation.  A multiple-spacecraft ``string-of-pearls'' or interleaved scan architecture reduces wall-clock time, enables repeated sampling of the same image-plane positions at controlled rotational phases, and provides redundancy for separating planet variability from instrument drift \cite{TuryshevEtAl2020NIAC,FriedmanEtAl2024,Helvajian2023}.  In an ideal interleaved architecture, $N_{\rm sc}$ independent spacecraft reduce the wall-clock dwell time approximately as $341/N_{\rm sc}\,\mathrm{d}$: for example, $N_{\rm sc}=4,8,$ and $16$ correspond to about $85$, $43$, and $21$ days of cumulative dwell per full raster before overhead.  The mission problem is therefore to combine propulsion to the $z\gtrsim548\,\AU$ focal region, continuous image-plane metrology, time-tagged photometry, communication of the measurement and calibration state, and dynamic reconstruction into a closed observing system.

\begin{figure}[h!]
\centering
\includegraphics[width=0.82\textwidth]{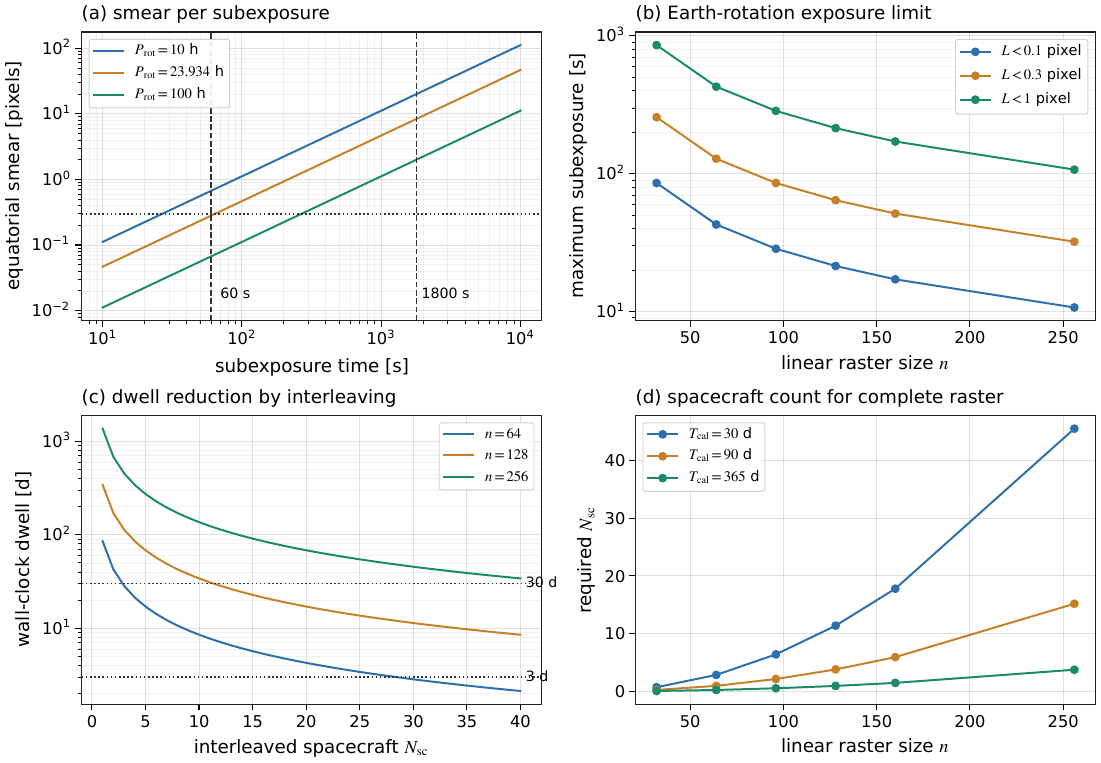}
\caption{Temporal sampling and architecture requirements computed from the benchmark raster geometry.  Panel (a) gives smear per subexposure for representative rotation periods using Eq.~\eqref{eq:smear}; the 60 s and 1800 s guides are horizontal text annotations.  Panel (b) gives the maximum Earth-rotation subexposure time as a function of grid size and tolerated smear.  Panel (c) shows how interleaved spacecraft reduce wall-clock dwell for different rasters when each image-plane sample accumulates $1800\,\s$ total dwell for one registered scalar map.  Panel (d) gives the spacecraft count required to complete one such raster in a specified calendar time.  Cloud-aware recovery multiplies this photon-time budget by the number of registered visits per phase bin, $M_p$, and by the number of retained phase bins, $N_\phi$, as in Eq.~\eqref{eq:cloud_calendar_budget}; the C3b/C3c demonstrations use $M_p=16$ and $N_\phi=1$.  The dwell values in this figure use the fiducial \(d=1\,{\rm m}\) photon budget; the explicit collector-diameter scaling of the cloud-aware photon-time budget is shown separately in Fig.~\ref{fig:diameter-cloud-scaling}. Temporal sampling is therefore part of the instrument design: the same interleaving that reduces wall-clock time also supplies the redundant phase coverage needed for cloud masking, statistical averaging, and dynamic reconstruction.}
\label{fig:temporal_architecture}
\end{figure}

Figure~\ref{fig:temporal_architecture} shows the corresponding subexposure, raster-size, and spacecraft-count scalings and emphasizes that cloud-aware recovery multiplies the one-map dwell by the number of registered visits and retained phase bins.

\begin{figure*}
    \centering
    \includegraphics[width=0.82\textwidth]{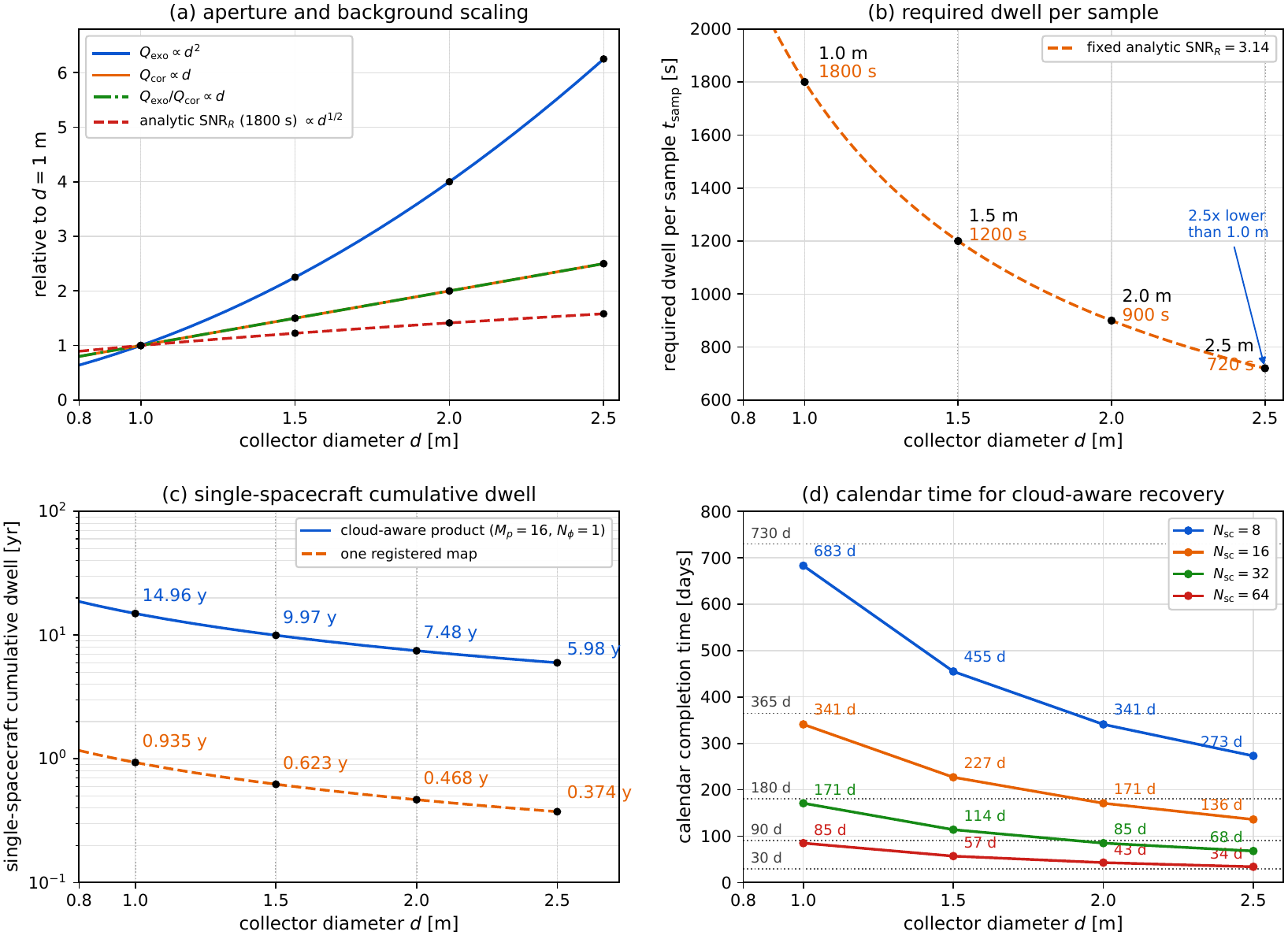}
\caption{
Collector-diameter scaling and cloud-aware observing-time trade for the scalar SGL benchmark. 
The curves are normalized to the fiducial \(d=1\,{\rm m}\), \(z_0=30\,{\rm pc}\), \(z=650\,{\rm AU}\), \(n=128\) case and use an aperture/FOV rescaling in which \(Q_{\rm exo}\propto d^2\) while the admitted coronal background scales as \(Q_{\rm cor}\propto d\), corresponding to an effective annular extraction width proportional to \(\lambda/d\). 
Panel (a) shows the resulting \(Q_{\rm exo}\), \(Q_{\rm cor}\), \(Q_{\rm exo}/Q_{\rm cor}\), and analytic fixed-dwell \(\SNRr\propto d^{1/2}\) scalings. Panel (b) shows the dwell per image-plane sample required to maintain the fiducial analytic reconstruction threshold \(\SNRr=3.14\), giving \(t_{\rm samp}=1800\,{\rm s}/[d/(1\,{\rm m})]\). 
Panel (c) converts this dwell into cumulative single-spacecraft time for one registered scalar map and for the \(M_p=16\), \(N_\phi=1\) cloud-aware product.  Panel (d) converts the same photon-time budget into calendar completion time for representative interleaved spacecraft counts.  The values exclude slew, calibration, communication, and scheduling overhead. The figure is an analytic observing-budget rescaling, not a new \(d\)-dependent image reconstruction; larger \(d\) reduces photon time under the stated extraction-width assumption, but it does not reduce the required number of independent cloud epochs, phase-registration accuracy, or the need for a dynamic cloud-aware inverse model.
}
 \label{fig:diameter-cloud-scaling}
\end{figure*}

A practical cloud-aware scan should not complete one raster and then attempt to deblur it as a static image.  It should cycle through image-plane positions and phase bins so that each surface longitude/latitude element is sampled repeatedly at comparable illumination, with subexposures short enough to keep $L_{\rm pix}\ll1$.  Samples or ring sectors predicted to be cloud dominated can be masked or downweighted, clear or low-cloud samples can be coadded after phase registration, and the remaining temporal signal can be reconstructed as a cloud-occurrence or variability map.  This observing pattern turns cloud contamination from an unmodeled bias into a controlled component of the data model, provided that the returned telemetry preserves the time, pointing, calibration, and background state of each sample.

Collector diameter enters this cloud-aware budget through both photon collection and the effective background admitted by the ring-extraction statistic. The fiducial photon rates in Eqs.~\eqref{eq:Qexo}--\eqref{eq:Qcor} define the \(d=1\,{\rm m}\) scalar count normalization. Figure~\ref{fig:diameter-cloud-scaling} applies a design rescaling about that normalization in which the collected exoplanet signal scales as \(Q_{\rm exo}\propto d^2\), while the admitted coronal shot-noise background scales as \(Q_{\rm cor}\propto d\) because the effective annular extraction width is taken to scale as \(\lambda/d\). In the corona-dominated limit this gives
\[
\SNRc \propto \frac{Q_{\rm exo}}{\sqrt{Q_{\rm cor}}}\,t_{\rm samp}^{1/2}
       \propto d^{3/2}t_{\rm samp}^{1/2}.
\]
Combining this with the analytic sampling penalty in Eq.~\eqref{eq:penalty}, \(\SNRr/\SNRc\propto d^{-1}\), gives \(\SNRr\propto d^{1/2}t_{\rm samp}^{1/2}\). 

Therefore, holding the fiducial analytic reconstruction threshold \(\SNRr=3.14\) fixed gives $t_{\rm samp}(d)=1800\,{\rm s}\,(1\,{\rm m}/d)$. The corresponding per-sample dwell values are \(1800\), \(1200\), \(900\), and \(720\,{\rm s}\) for \(d=1\), \(1.5\), \(2\), and \(2.5\,{\rm m}\), respectively. The cloud-aware \(M_p=16\), \(N_\phi=1\) photon-time budget then scales as \(14.95\,{\rm yr}/D\) before overhead. This is a photon-time and calendar-time rescaling, not a reduction in the number of statistically independent cloud realizations: \(M_p\) is still set by the residual cloud amplitude, cloud correlations, and the desired persistent-map uncertainty in Eqs.~\eqref{eq:cloud_epochs} and \eqref{eq:clear_samples}. Conversely, if the fixed-throughput scalar scaling \(Q_{\rm cor}\propto d^2\) in Eq.~\eqref{eq:Qcor} is retained without the \(\lambda/d\) extraction-width rescaling, the dwell-time curves in Fig.~\ref{fig:diameter-cloud-scaling} should be read as a design sensitivity rather than as the fiducial scalar count-model prediction. Thus Fig.~\ref{fig:diameter-cloud-scaling} separates a plausible aperture/FOV trade from the stronger claim, not made here, that larger aperture alone improves the cloud-recovered image metrics without a full \(d\)-dependent reconstruction.

Communications and autonomy enter the same architecture trade.  A scalar $128\times128$ raster contains $n^2=16384$ photometric samples per band and phase bin before subexposure, calibration, housekeeping, ring-sector, or spectral expansion.  The downlink from several hundred AU is therefore a mission-system requirement, but not an independent optical limitation: the measurement model defines which time-tagged photometric samples, calibration data, and reconstructed summary products must be returned for verification and reprocessing.

The occulting architecture is also critical.  Internal coronagraphs require a telescope large enough to separate the solar disk from the Einstein ring and impose throughput penalties.  External occulters relax the telescope diffraction constraint and can broaden the usable bandpass while reducing ring-extraction systematics \cite{TuryshevToth2022Spectral}.  For the imaging problem considered here, the first requirement is robust broadband mapping with a calibrated scalar ring statistic; spectroscopy, spectropolarimetry, and temporal monitoring are follow-on diagnostics that become scientifically useful after the map geometry, calibration, and dynamic inverse problem are under control.

\section{Comparison with other exoplanet-imaging approaches}
\label{sec:comparison}

The SGL is best compared to conventional facilities as an in-situ gravitational imaging system rather than as a large aperture: the spacecraft must be placed beyond $\sim547.8\,\AU$ and scanned across a moving image cylinder.  The comparison is decisive in one respect: conventional remote systems cannot reach the required angular resolution and photon budget simultaneously \cite{Turyshev2026exoimage}.

For a conventional coronagraph, starshade, or interferometer, a useful per-map-element exposure equation is
\begin{equation}
 \SNR_{\rm conv} =
 \frac{Q_{p,{\rm pix}}t}{\left[(Q_{p,{\rm pix}}+Q_{\rm zodi}+Q_{\rm exozodi}+Q_{\star,{\rm leak}}+Q_{\rm det})t+\sigma_{\rm speck}^2t^2\right]^{1/2}},
\label{eq:conventional_snr}
\end{equation}
This comparison is subject to the independent requirement that the facility actually resolve the planet with an aperture or baseline of order $10^2\,\km$.  For the SGL, the corresponding convolved-image SNR is Eq.~\eqref{eq:snrc}, followed by the deconvolution penalty in Eq.~\eqref{eq:penalty} and the support-restricted image metrics in Table~\ref{tab:metrics}.  The SGL advantage is thus not that every noise term disappears; rather, the angular-resolution term is supplied by the solar gravitational field and the remaining dominant terms become solar-corona shot noise, calibration, scan metrology, and dynamic inversion.

\begin{figure}[h]
\centering
\includegraphics[width=0.82\textwidth]{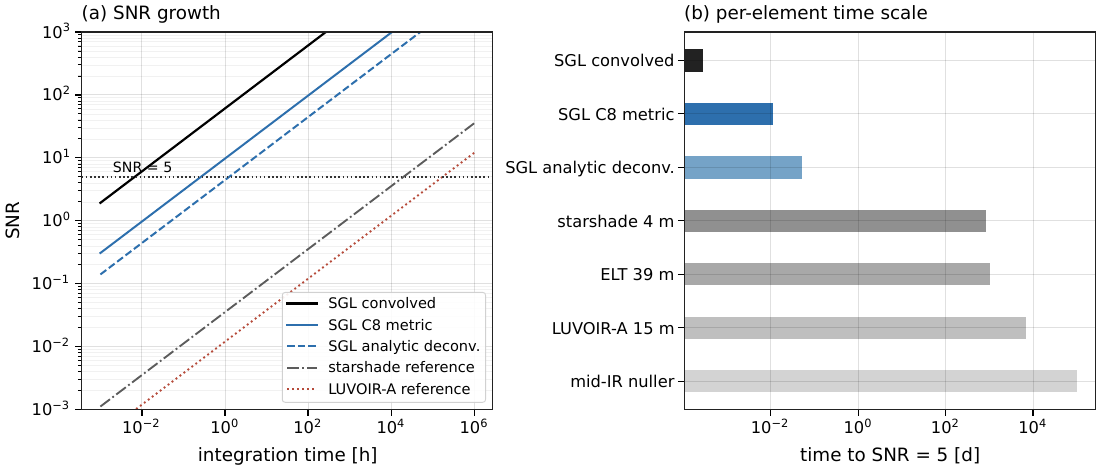}
\caption{SNR contrast between the SGL benchmark and representative conventional remote-imaging reference scalings.  Panel (a) compares SNR growth with integration time for the SGL convolved signal, the analytic deconvolution estimate, the C8 support-restricted metric, and two explicit conventional reference laws, $\SNR=5.9\times10^{-4}\sqrt{t/\s}$ and $\SNR=2.0\times10^{-4}\sqrt{t/\s}$, representing notional starshade and LUVOIR-A micro-pixels.  Panel (b) gives the resulting time to SNR $=5$: $2.80\times10^{-4}$ d (SGL convolved), $1.10\times10^{-2}$ d (C8 support-restricted metric), $5.28\times10^{-2}$ d (analytic deconvolution), $2.28$ yr (starshade 4 m), $2.84$ yr (ELT 39 m), $19.8$ yr (LUVOIR-A 15 m), and $272$ yr (mid-IR nuller).  All plotted SGL values are computed from the benchmark equations and parameters used in this paper; the conventional values are simple reference scalings used only to show the scale separation between an SGL image element and a conventional remote micro-pixel.}
\label{fig:sgl_vs_conventional_snr}
\end{figure}

The comparison in Fig.~\ref{fig:sgl_vs_conventional_snr} is computed as follows.  The SGL curves in panel~(a) use the fiducial benchmark parameters: $\SNRc(1800\,\s)=43.16$, the analytic deconvolution factor is $0.0728$, and the measured C8 ratio is $6.89/43.16=0.1596$.  The corresponding SGL times in panel~(b) use
\(t_5=t_0\left({5}/{\SNR_0}\right)^2\),  \(t_0=1800\,\s,\)
giving $2.80\times10^{-4}$ d for the convolved scalar signal, $1.10\times10^{-2}$ d for the C8 support-restricted metric, and $5.28\times10^{-2}$ d for the analytic deconvolution estimate.  The conventional curves are intentionally simple comparison anchors rather than mission forecasts: they use background-limited relations $\SNR=\alpha\sqrt{t}$ with $\alpha=5.9\times10^{-4}\,\sqrt{\s^{-1}}$ for a notional starshade micro-pixel and $\alpha=2.0\times10^{-4}\,\sqrt{\s^{-1}}$ for a LUVOIR-A-like micro-pixel.  The ELT and mid-infrared nuller bars are order-of-magnitude comparison anchors included only to emphasize that even optimistic conventional concepts remain limited primarily by the simultaneous aperture-or-baseline requirement and the very small planetary photon rate per resolved element \cite{Turyshev2026exoimage,LUVOIR2019,HabEx2020,RomanCGI2025,StarshadeRendezvous2019}.  Every curve or bar in Fig.~\ref{fig:sgl_vs_conventional_snr} is therefore either computed directly from the benchmark equations used in this paper or explicitly identified as a simplified comparison anchor.

A single filled aperture needs a diameter of order $160\,\km$ for a $10\times10$ Earth map at $10\,\pc$ in the visible.  A starshade suppresses starlight but does not improve the telescope's diffraction limit.  A space interferometer needs baselines of order $100\,\km$ and substantial collecting area while also maintaining path-length stability and high-contrast suppression.  Ground-based ELTs are photon-rich relative to small space telescopes but are diffraction-limited at milliarcsecond scales and face atmospheric residuals.  Indirect methods can infer longitudinal color maps or low-order surface components, but they do not produce resolved images at the surface scales considered here.  These limitations are derived quantitatively in the companion comparative exoplanet-imaging analysis and are consistent with mission studies for LUVOIR, HabEx, Roman CGI, starshade rendezvous, and interferometric concepts \cite{Turyshev2026exoimage,LUVOIR2019,HabEx2020,RomanCGI2025,StarshadeRendezvous2019,Bracewell1978,Angel1997}.

The SGL advantage is that the solar gravitational field supplies the angular response and optical gain that would otherwise require an aperture or interferometric baseline far beyond conventional mission scales.  Its disadvantages are mission-specific: reaching and operating in the focal region, suppressing solar light, subtracting coronal backgrounds, navigating the image plane, modeling solar multipoles, maintaining communications, and solving a dynamic inverse problem.  These are difficult requirements, but they are identifiable engineering, calibration, and inference requirements rather than undefined physical barriers.  Within the comparison assumptions used here and in Ref.~\cite{Turyshev2026exoimage}, no conventional remote architecture examined offers a comparable path to $\sim100\,\km$-class resolved maps at tens of parsecs; spectroscopy and spectropolarimetry could add diagnostic context, but the decisive SGL advantage is the ability to form a resolved image in the first place.

\section{Limitations and next steps}
\label{sec:limitations}

The stated scalar forward model defines the scope of the present benchmark and, more importantly, the traceable path to a mission-level simulator.  The corresponding extensions are well specified by the same operator notation.  First, the solar multipole branch should be replaced by numerical evaluation of the wave-optical integral with solar $J_2,J_4,\ldots$, spin-axis orientation, limb and oblateness constraints, and spacecraft-target geometry.  Second, the real-Earth scene should be extended from a single image to time-tagged Earth observations or a GCM coupled to radiative transfer, with wavelength-dependent clouds, ocean glint, atmospheric scattering, and thermal emission.  Third, the normalized host-star and exozodiacal templates should be replaced by physical leakage calculations using host magnitude, orbit geometry, and coronagraphic or external-occulter propagation.  Fourth, the photon-transfer detector model should be extended to include cosmic rays, persistence, interpixel capacitance, bad pixels, and detailed calibration covariance.  Fifth, coordinate-warp navigation should be replaced by closed-loop spacecraft dynamics and ephemeris estimation, coupled to the propulsion and station-keeping architecture.  Sixth, the static Wiener inverse with support constraints should be extended to the joint dynamic reconstruction in Eq.~\eqref{eq:dynamic_inverse}.  Seventh, the communications model should close the loop between time-tagged raw photometry, calibration data, onboard processing, and downlinked reconstruction products.

These extensions define the next simulation stage: (i) full multipole PSF libraries; (ii) synthetic multi-epoch Earth data or GCM scenes; (iii) wavelength-dependent ring photometry and starshade/coronagraph propagation; (iv) full detector calibration covariance; (v) closed-loop navigation and ephemeris models; and (vi) posterior uncertainty over the planet map and nuisance parameters.  The C3b and C3c branches added here are deliberately compact benchmark-level recovery tests.  C3b tests the mitigation path using advected diagnostic cloud-proxy epochs, while C3c removes one specific vulnerability by replacing that morphology with independent stochastic cloud fields matched only in low-order cloud statistics.  Quantitative validation of the full cloud-recovery estimator in Eqs.~\eqref{eq:phase_registered_data}--\eqref{eq:cloud_average} against physical multi-epoch Earth observations, GCM output, or radiative-transfer cloud fields remains a designated next-stage deliverable.  The present work supplies the baseline against which those upgrades can be measured.

It is useful to distinguish physical constraints from implementation requirements.  The broad aperture-averaged SGL response and the associated deconvolution penalty are optical properties of the lens.  Solar-corona shot noise is irreducible once coronal photons enter the measurement, even if the mean coronal brightness is accurately modeled and subtracted.  Planetary rotation and cloud evolution are astrophysical time-variability terms that must be included in the forward model.  By contrast, the accuracy of the solar-multipole PSF library, structured coronal residual templates, host-star leakage model, detector calibration covariance, image-plane metrology, scan strategy, ephemeris estimation, cloud-state prior, communications architecture, and inverse reconstruction are mission-design and implementation requirements.  The benchmark therefore does not make the SGL easy; it makes the feasibility question explicit.  The required observatory is demanding, but its controlling terms are quantitative and testable: propulsion to the focal region, calibrated ring extraction, solar-background suppression, sub-ppm effective calibration residuals, sub-meter image-plane metrology, physical PSF libraries, and dynamic reconstruction.

\section{Conclusions}
\label{sec:conclusions}

We have presented a scalar, aperture-averaged, real-Earth benchmark for imaging an Earth-like exoplanet with the Solar Gravitational Lens.  The benchmark is intentionally transparent: it uses an explicit SGL convolution operator, specified photon and calibration terms, controlled surrogates for several first-order systematics, and a stated regularized inverse.  Within that scope, the results make a technically defensible case that the SGL should be treated as an observatory architecture with quantitative requirements, not simply as a theoretical lens or a high-gain photometer.

First, the calibrated monopole SGL blur is not the dominant modeled limitation.  In the scalar aperture-integrated approximation, a known SGL kernel can be inverted.  The limiting terms are photon statistics, corona/detector calibration residuals, temporal sampling, navigation errors, and imperfect PSF knowledge.  The small C1a--C1b penalties have a specific interpretation: the adopted few-percent anisotropic tail perturbation is mild in this benchmark, while the true solar quadrupole, higher multipoles, and focal-plane caustic structure remain part of the mission-level PSF calibration problem.

Second, temporal modeling is intrinsic to SGL imaging because the data set is assembled from sequential, time-tagged image-plane samples.  The C3 branch shows that a static inverse applied to a rotating cloudy raster is the wrong estimator.  It should not be read as evidence that clouds defeat SGL imaging.  The C3b proxy branch gives the complementary quantitative result for advected diagnostic cloud-proxy epochs, while C3c tests the same mitigation path with cloud morphology generated independently of the surface proxy and independently across epochs.  Together these compact branches show that phase registration, cloud masking or downweighting, robust multi-epoch coaddition, and repeated phase coverage can recover persistent surface information in controlled scalar cloud-stress tests.  They do not constitute physical cloud validation.  The appropriate products are persistent or dynamic maps reconstructed from short subexposures, phase registration, repeated phase coverage, cloud masking or downweighting, statistical averaging, and time-dependent inversion.  Clouds are nuisance structure for a persistent surface map, but they are also atmospheric and climate observables in a dynamic SGL data set.

For the fiducial rotating-cloudy scalar stress tests, the stated C3b/C3c recovery metrics assume $M_p=16$ registered visits per phase bin, $t_{\rm samp}=1800\,\s$ per image-plane sample per visit, and $t_{\rm sub}=60\,\s$ subexposures; this corresponds to $14.95\,{\rm yr}$ of single-spacecraft cumulative dwell per registered phase bin before overhead, or approximately $341\,{\rm d}$ with $16$ interleaved spacecraft, with additional factors of $N_\phi$ for phase-resolved products. Under the aperture/FOV rescaling shown in Fig.~\ref{fig:diameter-cloud-scaling}, larger collector diameters reduce this photon-time component of the cloud-aware budget, but they do not remove the need for \(M_p\) independent registered visits, short subexposures, cloud masking or downweighting, and a dynamic inverse model.

Third, the calibration requirement is severe but quantifiable.  With $Q_{\rm cor}/Q_{\rm exo}=7.74\times10^4$, effective multiplicative residuals on the bright corona must be suppressed to sub-ppm levels after annular averaging, ring extraction, coronal subtraction, detector calibration, and background-template removal.  This is a benchmark-derived calibration requirement; demonstrating it requires the explicit ring-extraction, coronal-subtraction, detector-calibration, and background-template covariance model defined schematically by Eqs.~\eqref{eq:count_model}--\eqref{eq:effective_covariance}.

Fourth, under the scalar ring-extraction and mitigation assumptions stated in the benchmark, the mitigated scalar, phase-registered case at $30\,\pc$ recovers a scientifically recognizable spatial-contrast pattern at a $\sim200$--$230\,\km$ resolution proxy.  The C3b cloud-proxy branch recovers the surface-like proxy with $\SSIM=0.813$ and $\FRC_{50}=220\,\km$, and the C3c independent-cloud branch gives $\SSIM=0.899$ and a grid-limited $\FRC_{50}=199\,\km$, reducing the possibility that the cloud-recovery result is an artifact of a single advected cloud proxy.  Neither branch is a physical cloud retrieval, and the recognizable morphology should not be interpreted as proof that individual recovered features are physically real surface structures in a mission-level observation.  Increasing photon statistics to $\SNRc=80$ yields a higher-quality scalar benchmark map with $\SSIM=0.927$ and $\FRC_{50}=199\,\km$.  These quoted values are the performance of the stated scalar benchmark under its stated ring-extraction, calibration, temporal, navigation, and inversion assumptions.  Model-form uncertainty from the true solar multipole PSF, time-dependent cloud fields, host-star leakage geometry, full detector calibration covariance, ring-extraction optics, and closed-loop spacecraft navigation enters the mission-level forward-model extension identified above.

The strongest science cases are calibrated imaging products from repeated, time-tagged SGL observations: persistent surface and albedo maps, cloud and climate statistics, rotationally resolved surface--atmosphere structure, and temporal monitoring of major planetary regions.  Disk-integrated spectroscopy, spectropolarimetry, and rotational variability would add diagnostic context for habitability and biosignatures, but the decisive SGL advantage is the ability to form resolved maps at surface scales that are inaccessible to conventional remote telescopes or interferometers.

The SGL should therefore be treated as a technically plausible but demanding observatory architecture for resolved exoplanet imaging.  The present scalar benchmark shows that, when temporal sampling, solar-background suppression, detector calibration, image-plane metrology, PSF knowledge, calibrated ring extraction, communications, and regularized inversion meet the quantitative requirements identified above, recognizable $200$--$230\,\km$-class broadband spatial-contrast maps of an Earth-like planet at tens of parsecs are recoverable in the scalar model, with the lower end set by the two-pixel numerical floor of the adopted raster. The benchmark also identifies the forward-model ingredients that determine the next level of fidelity: a solar-multipole wave-optical PSF, explicit coronagraphic or external-occulter propagation for Einstein-ring extraction, physical coronal subtraction and background-control models, wavelength-dependent planet and atmosphere radiative transfer, physical host-star and exozodiacal leakage, time-dependent cloud fields, detector calibration covariance, closed-loop navigation, end-to-end communications, and a dynamic inverse problem that estimates nuisance parameters together with the planet map.  Thus the scientific return of an SGL mission is set by the joint performance of propulsion, optics, scan strategy, calibration, navigation, communications, and inference--a demanding but well-defined path to an imaging capability not otherwise available.

\section*{Acknowledgments} 
The work described here was carried out at the Jet Propulsion Laboratory, California Institute of Technology, Pasadena, California, under a contract with the National Aeronautics and Space Administration.

%

\end{document}